\documentclass[10pt,a4paper]{article}
\usepackage{amsmath, amssymb}
\usepackage{graphicx}
\usepackage{enumerate}
\usepackage[sort&compress,numbers]{natbib}
\usepackage[colorlinks,linkcolor=blue,anchorcolor=blue,citecolor=blue,dvipdfm]{hyperref}
\usepackage{subfigure}
\usepackage{colortbl}
\usepackage{tabularx}
\usepackage{indentfirst}
\usepackage{multirow}
\usepackage{lscape}
\usepackage{color}

\usepackage[hmargin={2.5cm,2.5cm}, vmargin={2.5cm,2.5cm}]{geometry}
\renewcommand{\baselinestretch}{1.28}

\begin{document}

\title{\Large \textbf{Bidirectional Control of Absence Seizures by the Basal Ganglia: A Computational Evidence}}
\author{Mingming Chen$^{1,}$\thanks{These authors contributed to this work equally} , Daqing Guo$^{1,}$\footnotemark[1]~$^{,}$\thanks{Corresponding authors: DY: dyao@uestc.edu.cn; DG: dqguo@uestc.edu.cn. Tel: +86-028-83201018, Fax: +86-028-83208238, Address: $\#$4, Section 2, North JianShe Road, Chengdu 610054, People's Republic of China.} , Tiebin Wang$^{1}$, Wei Jing$^{1}$, Yang Xia$^{1}$, \\Peng Xu$^{1}$, Cheng Luo$^{1}$, Pedro A. Valdes-Sosa$^{1, 2}$, and Dezhong Yao$^{1,}$\footnotemark[2]}
\date{\small $^1$Key Laboratory for NeuroInformation of Ministry of Education, School of Life Science and Technology, University of Electronic Science and Technology of China, Chengdu 610054, People's Republic of China\\ $^2$Cuban Neuroscience Center, Ave 25 $\#$15202 esquina 158, Cubanacan, Playa, Cuba\\ \today}

\maketitle
\begin{abstract}
Absence epilepsy is believed to be associated with the abnormal interactions between the cerebral cortex and thalamus. Besides the direct coupling, anatomical evidence indicates that the cerebral cortex and thalamus also communicate indirectly through an important intermediate bridge--basal ganglia. It has been thus postulated that the basal ganglia might play key roles in the modulation of absence seizures, but the relevant biophysical mechanisms are still not completely established. Using a biophysically based model, we demonstrate here that the typical absence seizure activities can be controlled and modulated by the direct GABAergic projections from the substantia nigra pars reticulata (SNr) to either the thalamic reticular nucleus (TRN) or the specific relay nuclei (SRN) of thalamus, through different biophysical mechanisms. Under certain conditions, these two types of seizure control are observed to coexist in the same network. More importantly, due to the competition between the inhibitory SNr-TRN and SNr-SRN pathways, we find that both decreasing and increasing the activation of SNr neurons from the normal level may considerably suppress the generation of SWDs in the coexistence region. Overall, these results highlight the bidirectional functional roles of basal ganglia in controlling and modulating absence seizures, and might provide novel insights into the therapeutic treatments of this brain disorder.
\end{abstract}


\section*{Author Summary}
Epilepsy is a general term for conditions with recurring seizures. Absence seizures are one of several kinds of seizures, which are characterized by typical 2-4 Hz spike-and-slow wave discharges (SWDs). There is accumulating evidence that absence seizures are due to abnormal interactions between cerebral cortex and thalamus, and the basal ganglia may take part in controlling such brain disease via the indirect basal ganglia-thalamic pathway relaying at superior colliculus. Actually, the basal ganglia not only send indirect signals to thalamus, but also communicate with several key nuclei of thalamus through multiple direct GABAergic projections. Nevertheless, whether and how these direct pathways regulate absence seizure activities are still remain unknown. By computational modelling, we predicted that two direct inhibitory basal ganglia-thalamic pathways emitting from the substantia nigra pars reticulata may also participate in the control of absence seizures. Furthermore, we showed that these two types of seizure control can coexist in the same network, and depending on the instant network state, both lowing and increasing the activation of SNr neurons may inhibit the SWDs due to the existence of competition. Our findings emphasize the bidirectional modulation effects of basal ganglia on absence seizures, and might have physiological implications on the treatment of absence epilepsy.

\section*{Introduction}
Absence epilepsy is a generalized non-convulsive seizure disorder of the brain, mainly occurring in the childhood years \cite{panayiotopoulos1997absence}. A typical attack of absence seizures is characterized by a brief loss of consciousness that starts and terminates abruptly, and meanwhile an electrophysiological hallmark, i.e. the bilaterally synchronous spike and wave discharges (SWDs) with a slow frequency at approximately 2-4 Hz, can be observed on the electroencephalogram (EEG) of patients \cite{panayiotopoulos1997absence, crunelli2002childhood}. There is a broad consensus that the generation of SWDs during absence seizures is due to the abnormal interactions between cerebral cortex and thalamus, which together form the so-called corticothalamic system. The direct evidence in support of this view is based on simultaneous recordings of cortex and thalamus from both rodent animal models and clinical patients \cite{marescaux1995genetic, coenen2003genetic, timofeev2004neocortical}. Recent computational modelling studies on this prominent brain disorder also approved the above viewpoint and provided more deep insights into the possible generation mechanism of SWDs in the corticothalamic system \cite{destexhe1998spike, destexhe1999can, robinson2002dynamics, lopes2003epilepsies, suffczynski2004dynamics, breakspear2006unifying, lytton2008computer, marten2009onset}.

The basal ganglia comprise a group of interconnected subcortical nucleus and, as a whole, represent one fundamental processing unit of the brain. It has been reported that the basal ganglia are highly associated with a variety of brain functions and diseases, such as cognitive \cite{stocco2010conditional}, emotional functions \cite{packard2002learning}, motor control \cite{groenewegen2003basal}, Parkinson's disease \cite{gatev2009interactions, Kumar2011BG}, and epilepsy \cite{paz2005rhythmic, luo2012resting}. Anatomically, the basal ganglia receive multiple projections from both the cerebral cortex and thalamus, and in turn send both direct and indirect output projections to the thalamus. These connections enable the activities of the basal ganglia to influence the dynamics of the corticothalamic system. Therefore, it is naturally expected that the basal ganglia may provide an active role in mediating between seizure and non-seizure states for absence epileptic patients. Such hypothesis has been confirmed by both previous animal experiments \cite{deransart1998role, deransart2002control, paz2005rhythmic, paz2007activity} and recent human neuroimage data \cite{luo2012resting, biraben2004pet, postuma2006basal}. Nevertheless, due to the complicated interactions between basal ganglia and thalamus, the underlying neural mechanisms on how the basal ganglia control the absence seizure activities are still remain unclear.

From the anatomical perspective, the substantia nigra pars reticulata (SNr) is one of the major output nucleus of the basal ganglia to thalamus. Previous experimental studies using various rodent animal models have demonstrated that suitable changes in the firing of SNr neurons can modulate the occurrence of absence seizures \cite{deransart1998role, deransart2002control, paz2007activity, Kase2012}. Specifically, it has been found that pharmacological inactivation of the SNr by injecting ${\gamma}$-aminobutyric acids (GABA) agonists or glutamate antagonists suppresses absence seizures\cite{deransart1998role, deransart2002control}. Such antiepileptic effect was supposed to be attributed to the overall inhibitory effect of the indirect pathway from the SNr to thalamic reticular nucleus (TRN) relaying at superior colliculus \cite{deransart1998role, deransart2002control}. In addition to this indirect inhibitory pathway, it is known that the SNr also contains GABAergic neurons directly projecting to the TRN and specific relay nuclei (SRN) of thalamus \cite{Haber2009, Gulcebi2012}. Theoretically, changing the activation level of SNr may also significantly impact the firing activities of SRN and TRN neurons \cite{Gulcebi2012, Chiara19}. This contribution might further interrupt the occurrence of SWDs in the corticothalamic system, thus providing an alternative mechanism to regulate typical absence seizure activities. To our knowledge, however, so far the precise roles of these direct basal ganglia-thalamic pathways in controlling absence seizures are not completely established.

To address this question, we develop a realistic mean-field model for the basal ganglia-corticothalamic (BGCT) network in the present study. Using  various dynamic analysis techniques, we show that the absence seizures are controlled and modulated either by the isolated SNr-TRN pathway or the isolated SNr-SRN pathway. Under suitable conditions, these two types of modulations are observed to coexist in the same network. Importantly, in this coexist region, both low and high activation levels of SNr neurons can suppress the occurrence of SWDs due to the competition between these two direct inhibitory basal ganglia-thalamic pathways. These findings clearly outline a bidirectional control of absence seizures by the basal ganglia, which is a novel phenomenon that has never been identified both in previous experimental and modelling studies. Our results, on the one hand, further improve the understanding of the significant role of basal ganglia in controlling absence seizure activities, and on the other hand, provide testable hypotheses for future experimental studies.

\section*{Materials and Methods}
\subsection*{Model}

\begin{figure*}[tp]
\center
\includegraphics[width=9cm]{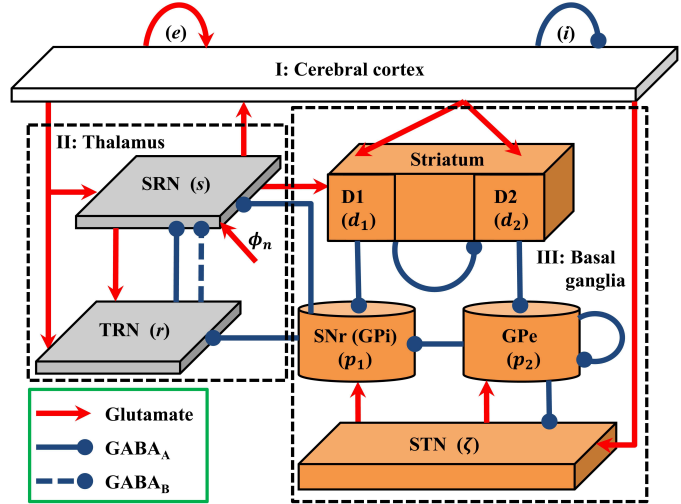}
\caption{\textbf{Framework of the basal ganglia-corticothalamic network.} Neural populations include $e=$ excitatory pyramidal neurons; $i=$ inhibitory interneurons; $r=$ thalamic reticular nucleus (TRN); $s=$ specific relay nuclei (SRN); $d_1=$ striatal D1 neurons; $d_2=$ striatal D2 neurons; $p_1=$ substantia nigra pars reticulata (SNr); $p_2=$ globus pallidus external (GPe) segment; $\zeta=$ subthalamic nucleus (STN). Note that we do not model the globus pallidus internal (GPi) segment independently but consider SNr and GPi as a signal structure in this work. Red lines with arrow heads denote the excitatory projections mediated by glutamate receptors. Blue solid and dashed lines with round heads represent the inhibitory projections mediated by GABA$_{\text{A}}$ and GABA$_{\text{B}}$ receptors, respectively. Compared with the traditional model of corticothalamic system on absence seizures, the basal ganglia are also included in our biophysical model.}
\label{fig:Fig.1}
\end{figure*}

We build a biophysically based model that describes the population dynamics of the BGCT network to investigate the possible roles of basal ganglia in the control of absence seizures. The network framework of this model is inspired by recent modelling studies on Parkinson's disease \cite{van2009amean, van2009bmean}, which is shown schematically in Fig.~1. The network totally includes nine neural populations, which are indicated as follows: $e=$ excitatory pyramidal neurons; $i=$ inhibitory interneurons; $r=$ TRN; $s=$ SRN; $d_1=$ striatal D1 neurons; $d_2=$ striatal D2 neurons; $p_1=$ SNr; $p_2=$ globus pallidus external (GPe) segment; $\zeta=$ subthalamic nucleus (STN). Similar to other modelling studies \cite{van2009amean, van2009bmean, BarGad2003}, we do not model the globus pallidus internal (GPi) segment independently but consider SNr and GPi as a signal structure in the present study, because they are reported to have closely related inputs and outputs, as well as similarities in cytology and function. Three types of neural projections are contained in the BGCT network. For sake of clarity, we employ different line types and heads to distinguish them (see Fig.~1). The red lines with arrow heads denote the excitatory projections mediated by glutamate, whereas the blue solid and dashed lines with round heads represent the inhibitory projections mediated by ${\text{GABA}_{\text{A}}}$ and ${\text{GABA}_{\text{B}}}$, respectively. It should be noted that in the present study the connections among different neural populations are mainly inspired by previous modelling studies \cite{van2009amean, van2009bmean}. Additionally, we also add the connection sending from SNr to TRN in our model, because recent anatomical findings have provided evidence that the SNr also contains GABAergic neurons directly projecting to the TRN \cite{Haber2009, Gulcebi2012, Chiara19}.

The dynamics of neural populations are characterized by the mean-field model \cite{robinson1997propagation, PhysRevE.63.021903, robinson2002dynamics, breakspear2006unifying, PhysRevE.58.3557,robinson2002dynamics}, which was proposed to study the macroscopic dynamics of neural populations in a simple yet efficient way. The first component of the mean-field model describes the average response of populations of neurons to changes in cell body potential. For each neural population, the relationship between the mean firing rate ${Q}_{a}$ and its corresponding mean membrane potential $V_{a}$ satisfies an increasing sigmoid function, given by
\begin{equation}
\begin{aligned}
Q_{a}(\textbf{r},t)\equiv{F}[{V}_{a}(\textbf{r},t)]=\frac{{Q}_{a}^{max}}{1+{\text{exp}}[-{\frac{\pi}{\sqrt{3}}}
{\frac{({V}_{a}(\textbf{r},t)-{\theta}_{a})}{\sigma}}]},
\label{eq:eq1}
\end{aligned}
\end{equation}
where $a\in A= \{e, i, r, s, d_1, d_2, p_1, p_2, \zeta\}$ indicate different neural populations, $Q_{a}^{max}$ denotes the maximum firing rate, \textbf{r} represents the spatial position, ${\theta}_{a}$ is the mean firing threshold, and ${\sigma}$ is the threshold variability of firing rate. If ${V}_{a}$ exceeds the threshold ${\theta}_{a}$, the neural population fires action potentials with an average firing rate $Q_{a}$. It should be noted that the sigmoid shape of $Q_{a}$ is physiologically crucial for this model, ensuring that the average firing rate cannot exceed the maximum firing rate $Q_{a}^{max}$. The changes of the average membrane potential $V_{a}$ at the position \textbf{r}, under incoming postsynaptic potentials from other neurons, are modeled as \cite{robinson1997propagation, PhysRevE.63.021903, HBM.HBM20032, PhysRevE.58.3557,robinson2002dynamics}
{\begin{equation}
\begin{aligned}
D_{{\alpha}{\beta}}{V}_{a}(\textbf{r},t)=\sum_{b \in A}{v}_{ab}\cdot{\phi_{b}}(\textbf{r},t),
\label{eq:eq2}
\end{aligned}
\end{equation}
\begin{equation}
\begin{aligned}
D_{{\alpha}{\beta}}=\frac{1}{{\alpha}{\beta}}[\frac{{\partial}^2}{\partial{t}^2}+
({\alpha}+{\beta}){\frac{\partial}{\partial{t}}}+{\alpha}{\beta}],
\label{eq:eq3}
\end{aligned}
\end{equation}}
where $D_{{\alpha}{\beta}}$ is a differential operator representing the dendritic filtering of incoming signals. $\alpha$ and $\beta$ are the decay and rise times of cell-body response to incoming signals, respectively. ${v}_{ab}$ is the coupling strength between neural populations of type ${a}$ and type ${b}$. ${\phi_{b}}(\textbf{r},t)$ is the incoming pulse rate from the neural population of type ${b}$ to type ${a}$. For simplicity, we do not consider the transmission delay among most neural populations in the present work. However, since the ${\text{GABA}}_{\text{B}}$ functions via second messenger processes, a delay parameter $\tau$ is introduced to its incoming pulse rate (i.e., ${\phi_{b}}(\textbf{r},t-\tau)$) to mimic its slow synaptic kinetics. This results in a delay differential equation in the final mathematical description of the BGCT model. Note that the similar modelling method has also been used in several previous studies \cite{marten2009onset, Destexhebook2001}.

In our system, each neural population gives rise to a field $\phi_a$ of pulses, which travels to other neural population at a mean conduction velocity $v_{a}$. In the continuum limit, this type of propagation can be well-approximated by a damped wave equation \cite{jirsa1996field, robinson1997propagation, PhysRevE.63.021903, PhysRevE.58.3557,robinson2002dynamics}:
\begin{equation}
\begin{aligned}
\frac{1}{\gamma_{a}^2}[{\frac{{\partial}^2}{\partial{t}^2}}+2\gamma_{a}\frac{{\partial}}{\partial{t}}
+{\gamma_{a}^2}-v_{a}^2{\nabla}^2]\phi_a(\textbf{r},t)=Q_a({\textbf{r},t}).
\label{eq:eq4}
\end{aligned}
\end{equation}
Here ${\nabla}^2$ is the Laplacian operator (the second spatial derivative), $r_{a}$ is the characteristic range of axons of type $a$, and $\gamma_{a}={v_{a}}/{r_{a}}$ governs the temporal damping rate of pulses. In our model, only the axons of cortical excitatory pyramidal neurons are assumed to be sufficiently long to yield significant propagation effect. For other neural populations, their axons are too short to support wave propagation on the relevant scales. This gives $\phi_{c}=F(V_c)$ ($c=i,r,s,d_1, d_2, p_1, p_2,\zeta$). Moreover, as one of typical generalized seizures, the dynamical activities of absence seizures are believed to occur simultaneously throughout the brain. A reasonable simplification is therefore to assume that the spatial activities are uniform in our model, which has been shown as the least stable mode in models of this class \cite{PhysRevE.63.021903, HBM.HBM20032,robinson1997propagation}. To this end, we ignore the spatial derivative and set ${\nabla}^2=0$ in Eq.~(\ref{eq:eq4}). Accordingly, the propagation effect of cortical excitatory axonal field $\phi_e$ is finally given by \cite{PhysRevE.63.021903, HBM.HBM20032,robinson1997propagation}:
\begin{equation}
\begin{aligned}
\frac{1}{\gamma_{e}^2}[{\frac{\text{d}^2}{\text{d}t^2}}+2\gamma_{e}{\frac{\text{d}}{\text{d}t}}
+{\gamma_{e}^2}]\phi_e(t)=Q_e(t),
\label{eq:eq5}
\end{aligned}
\end{equation}
where $\gamma_{e}={v_{e}}/{r_{e}}$. For the population of cortical inhibitory interneurons, the BGCT model can be further reduced by using $V_i=V_e$ and $Q_i=Q_e$, which is based on the assumption that intracortical connectivities are proportional to the numbers of synapses involved \cite{robinson1997propagation, PhysRevE.63.021903, HBM.HBM20032, marten2009onset,PhysRevE.58.3557,robinson2002dynamics}. It has been demonstrated that by making these above reductions, the developed BGCT model becomes computationally more tractable without significant deteriorating the precision of numerical results.

We then rewrite above equations in the first-order form for all neural populations. Following above assumptions, we use Eqs.~(\ref{eq:eq1})-(\ref{eq:eq3}) and (\ref{eq:eq5}) for modelling the dynamics of excitatory pyramidal neurons, and Eqs.~(\ref{eq:eq1})-(\ref{eq:eq3}) for modelling the dynamics of other neural populations. This yields the final mathematical description of the BGCT model given as follows:
\begin{equation}
\begin{aligned}
\frac{\text{d}{\phi_e}(t)}{\text{d}t}=\dot{\phi}_{e}(t),
\end{aligned}
\label{eq:6}
\end{equation}
\begin{equation}
\begin{aligned}
\frac{\text{d}\dot{\phi}_{e}(t)}{\text{d}t}=\gamma_{e}^{2}[-{\phi_e(t)}+
{F(V_e(t))}]-2{\gamma_e}{\dot{\phi}_{e}(t)},
\end{aligned}
\label{eq:7}
\end{equation}
\begin{equation}
\frac{\text{d}X(t)}{\text{d}t}=\dot{X}(t),
\label{eq:8}
\end{equation}
\begin{equation}
\frac{\text{d}\dot{X}(t)}{\text{d}t}=CY(t)-W(t),
\label{eq:9}
\end{equation}
where
\begin{equation}
X(t)=\left[ \begin{array}{c}
  V_e(t),V_{d_1}(t),V_{d_2}(t),V_{p_1}(t),V_{p_2}(t),V_{\zeta}(t),V_r(t),V_s(t)
\end{array}
\right]^{\rm T},
\label{eq:10}
\end{equation}
\begin{equation}
C={\alpha}{\beta},
\label{eq:11}
\end{equation}
\begin{equation}
W(t)=(\alpha+\beta)\dot{X}(t).
\label{eq:12}
\end{equation}
In Eq.~(\ref{eq:10}), the superscript T denotes transposition. The detailed expression of $Y(t)$ for different neural populations is represented by $C_1$, $Y_1(t)$ and $Y_2(t)$, given by
\begin{equation}
\begin{aligned}
Y(t)=C_{1}Y_{1}(t)-Y_{2}(t),
\end{aligned}
\label{eq:13}
\end{equation}
with
\begin{equation}
C_1=\left[\begin{array}{cccccccccc}
  v_{ee}&v_{ei}&0&0&0&0&0&(0,0)&v_{es}\\
  v_{{d_1}e}&0&v_{{d_1}{d_1}}&0&0&0&0&(0,0)&v_{{d_1}s} \\
  v_{{d_2}e}&0&0&v_{{d_2}{d_2}}&0&0&0&(0,0)&v_{{d_2}s}\\
  0&0&v_{{p_1}{d_1}}&0&0&v_{{p_1}{p_2}}&v_{{p_1}{\zeta}}&(0,0)&0\\
  0&0&0&v_{{p_2}{d_2}}&0&v_{{p_2}{p_2}}&v_{{p_2}{\zeta}}&(0,0)&0\\
 v_{{\zeta}e}&0&0&0&0&v_{{\zeta}{p_2}}&0&(0,0)&0\\
 v_{re}&0&0&0&v_{rp_1}&0&0&(0,0)&v_{rs}\\
  v_{se}&0&0&0&v_{sp_1}&0&0&(v_{sr}^{A},v_{sr}^{B})&0
\end{array}
\right],
\label{eq:14}
\end{equation}
\begin{equation}
Y_1(t)=\left[\begin{array}{c}
  \phi_e,F(V_e),F(V_{d_1}),F(V_{d_2}),F(V_{p_1}),F(V_{p_2}),F(V_{\zeta}),\left( \begin{array}{cc}
  F(V_r),F(V_r(t-\tau))
    \end{array}
      \right),F(V_{s})
\end{array}
\right]^{\rm T},
\label{eq:15}
\end{equation}
\begin{equation}
Y_2(t)=\left[\begin{array}{c}
  V_e(t),V_{d_1}(t),V_{d_2}(t),V_{p_1}(t),V_{p_2}(t),V_{\zeta}(t),V_r(t),V_s(t)-\phi_n
\end{array}
\right]^{\rm T}.
\label{eq:16}
\end{equation}
Here the variable $\tau$ in Eq.~(\ref{eq:15}) denotes the ${\text{GABA}_{\text{B}}}$ delay and the parameter $\phi_n$ in Eq.~(\ref{eq:16}) represents the constant nonspecific subthalamic input onto SRN.

The parameters used in our BGCT model are compatible with physiological experiments and their values are adapted from previous studies \cite{breakspear2006unifying, marten2009onset, robinson2002dynamics, van2009amean, van2009bmean, HBM.HBM20032}. Unless otherwise noted, we use the default parameter values listed in Table 1 for numerical simulations. Most of the default values of these parameters given in Table 1 are based on either their nominal values or parameter ranges reported in above literature. A small number of parameters associated with the basal ganglia (i.e., $v_{d_1d_1}$, $v_{p_2p_2}$ and $v_{{p_2}{\zeta}}$) are adjusted slightly, but still within their normal physiological ranges, to ensure our developed model can generate the stable 2-4 Hz SWDs under certain conditions. Note that due to lack of quantitative data, the coupling strength of the SNr-TRN pathway needs to be estimated. Considering that the SNr sends GABAergic projections both to SRN and TRN and also both of these two nuclei are involved in thalamus, it is reasonable to infer that the coupling strengths of these two pathways are comparable. For simplicity, here we chose $v_{rp_1}=v_{sp_1}=-0.035$ mV~s by default. In the following studies, we also change (decrease or increase) the value of $v_{rp_1}$ several folds by employing a scale factor $K$ (see below) to examine how the inhibition from the SNr-TRN pathway regulates absence seizures. Additionally, during this study, several other critical parameters (i.e., $v_{sr}$, $\tau$ and $v_{p_1\zeta}$) are also varied within certain ranges to obtain different dynamical states and investigate their possible effects on the modulation of absence seizures.

\begin{table*}[tp]
\renewcommand{\arraystretch}{0.86}\small
\begin{tabular}{|l|l|l|l|l|}
\hline
\multicolumn{5}{|l|}{\bf{A: Maximum firing rate}}  \\  \hline
\bf{Symbol}& \multicolumn{2}{l|}{\bf{Description}}&\bf{Value}&\bf{References}\\ \hline
{$Q_{e}^{max}$, $Q_{i}^{max}$} &  \multicolumn{2}{l|}{Cortical maximum firing rate} & 250 $\text{Hz}$ & \cite{breakspear2006unifying,robinson2002dynamics,marten2009onset}\\ \hline
{$Q_{d_1}^{max}$, $Q_{d_2}^{max}$} &  \multicolumn{2}{l|}{Striatum maximum firing rate} & 65 $\text{Hz}$ & \cite{van2009amean,van2009bmean}\\ \hline
$Q_{p_1}^{max}$ &  \multicolumn{2}{l|}{SNr maximum firing rate} & 250 $\text{Hz}$ & \cite{van2009amean,van2009bmean}\\ \hline
$Q_{p_2}^{max}$ &  \multicolumn{2}{l|}{GPe maximum firing rate} & 300 $\text{Hz}$ & \cite{van2009amean,van2009bmean}\\ \hline
$Q_{\zeta}^{max}$ &  \multicolumn{2}{l|}{STN maximum firing rate} & 500 $\text{Hz}$ & \cite{van2009amean,van2009bmean}\\ \hline
$Q_{s}^{max}$ &  \multicolumn{2}{l|}{SRN maximum firing rate} & 250 $\text{Hz}$ & \cite{breakspear2006unifying,robinson2002dynamics,marten2009onset}\\ \hline
$Q_{r}^{max}$ &  \multicolumn{2}{l|}{TRN maximum firing rate} & 250 $\text{Hz}$ & \cite{breakspear2006unifying,robinson2002dynamics,marten2009onset}\\ \hline
\multicolumn{5}{|l|}{\bf{B: Mean firing threshold}}  \\  \hline
\bf{Symbol} & \multicolumn{2}{l|}{\bf{Description}} & \bf{Value} & \bf{References}\\ \hline
{$\theta_{e}$, $\theta_{i}$} & \multicolumn{2}{l|}{Mean firing threshold of cortical populations} & $15$ ${\text{mV}}$ &  \cite{breakspear2006unifying,robinson2002dynamics,marten2009onset,HBM.HBM20032}\\ \hline
{$\theta_{d_1}$, $\theta_{d_2}$} & \multicolumn{2}{l|}{Mean firing threshold of striatum} &$19$ $ {\text{mV}}$ & \cite{van2009amean,van2009bmean}\\ \hline
$\theta_{p_1}$ & \multicolumn{2}{l|}{Mean firing threshold of SNr} & $10$ ${\text{mV}}$ & \cite{van2009amean,van2009bmean}\\ \hline
$\theta_{p_2}$ & \multicolumn{2}{l|}{Mean firing threshold of GPe} & $9$ ${\text{mV}}$ & \cite{van2009amean,van2009bmean}\\ \hline
$\theta_{\zeta}$ & \multicolumn{2}{l|}{Mean firing threshold of STN} & $10$ ${\text{mV}}$ & \cite{van2009amean,van2009bmean}\\ \hline
$\theta_{s}$ & \multicolumn{2}{l|}{Mean firing threshold of SRN} & $15$ ${\text{mV}}$ &\cite{breakspear2006unifying,robinson2002dynamics,marten2009onset,HBM.HBM20032}\\ \hline
$\theta_{r}$ & \multicolumn{2}{l|}{Mean firing threshold of TRN} & $15$ ${\text{mV}}$ &\cite{breakspear2006unifying,robinson2002dynamics,marten2009onset,HBM.HBM20032}\\ \hline
\multicolumn{5}{|l|}{\bf{C: Coupling strength}}  \\  \hline
\bf{Symbol}&\bf{Source}&\bf{Target}&\bf{Value}&\bf{References}\\ \hline
$v_{ee}$ &  Excitatory pyramidal neurons &   Excitatory pyramidal neurons & $1$ ${\text{mV s}}$ & \cite{breakspear2006unifying,marten2009onset}\\ \hline
$-v_{ei}$ &  Inhibitory interneurons &  Excitatory pyramidal neurons & $1.8$ ${\text{mV s}}$ & \cite{breakspear2006unifying,marten2009onset}\\ \hline
$v_{re}$ &   Excitatory pyramidal neurons & TRN & $0.05$ ${\text{mV s}}$ & \cite{robinson2002dynamics,marten2009onset}\\ \hline
$v_{rs}$ &  SRN & TRN & $0.5$ ${\text{mV s}}$ & \cite{robinson2002dynamics,marten2009onset}\\ \hline
$-v_{sr}^{A,B}$ & TRN & SRN & $0.4-2$ ${\text{mV s}}$ &\cite{robinson2002dynamics,HBM.HBM20032}\\ \hline
$v_{{d_1}e}$ &  Excitatory pyramidal neurons & Striatal D1 neurons & $1$ ${\text{mV s}}$ & \cite{van2009amean,van2009bmean}\\ \hline
$-v_{{d_1}{d_1}}$ & Striatal D1 neurons & Striatal D1 neurons & $0.2$ ${\text{mV s}}$ & \cite{van2009amean,van2009bmean}\\ \hline
$v_{{d_1}s}$ & SRN & Striatal D1 neurons  & $0.1$ ${\text{mV s}}$ & \cite{van2009amean,van2009bmean}\\ \hline
$v_{{d_2}e}$ &  Excitatory pyramidal neurons & Striatal D2 neurons & $0.7$ ${\text{mV s}}$ & \cite{van2009amean,van2009bmean}\\ \hline
$-v_{{d_2}{d_2}}$ & Striatal D2 neurons & Striatal D2 neurons & $0.3$ ${\text{mV s}}$ & \cite{van2009amean,van2009bmean}\\ \hline
$v_{{d_2}s}$ & SRN & Striatal D2 neurons & $0.05$ ${\text{mV s}}$ & \cite{van2009amean,van2009bmean}\\ \hline
$-v_{{p_1}{d_1}}$ & Striatal D1 neurons & SNr & $0.1$ ${\text{mV s}}$ & \cite{van2009amean,van2009bmean}\\ \hline
$-v_{{p_1}{p_2}}$ & GPe & SNr &  $0.03$ ${\text{mV s}}$ & \cite{van2009amean,van2009bmean}\\ \hline
$v_{{p_1}{\zeta}}$ & STN & SNr &  $0-0.6$ ${\text{mV s}}$ & \cite{van2009amean,van2009bmean}\\ \hline
$-v_{{p_2}{d_2}}$ & Striatal D2 neurons & GPe & $0.3$ ${\text{mV s}}$ & \cite{van2009amean,van2009bmean}\\ \hline
$-v_{{p_2}{p_2}}$ &  GPe & GPe &  $0.075$ ${\text{mV s}}$ & \cite{van2009amean,van2009bmean}\\ \hline
$v_{{p_2}{\zeta}}$ & STN & GPe & $0.45$ ${\text{mV s}}$ & \cite{van2009amean,van2009bmean}\\ \hline
$-v_{{\zeta}{p_2}}$ & GPe & STN &  $0.04$ ${\text{mV s}}$ & \cite{van2009amean,van2009bmean}\\ \hline
$v_{es}$ & SRN &  Excitatory pyramidal neurons & $1.8$ ${\text{mV s}}$ & \cite{robinson2002dynamics,marten2009onset}\\ \hline
$v_{se}$ &  Excitatory pyramidal neurons & SRN & $2.2$ ${\text{mV s}}$ &\cite{robinson2002dynamics}\\ \hline
$v_{{\zeta}e}$ &  Excitatory pyramidal neurons & STN & $0.1$ ${\text{mV s}}$ & \cite{van2009amean,van2009bmean}\\ \hline
$-v_{s{p_1}}$ & SNr & SRN &  $0.035$ ${\text{mV s}}$ &  \cite{van2009amean,van2009bmean}\\ \hline
$-v_{r{p_1}}$ & SNr & TRN & $0.035$ ${\text{mV s}}$ & Estimated\\ \hline
\multicolumn{5}{|l|}{\bf{D: Other parameters}}  \\  \hline
\bf{Symbol}& \multicolumn{2}{l|}{\bf{Description}}&\bf{Value}&\bf{References}\\ \hline
$\gamma_{e}$ & \multicolumn{2}{l|}{Cortical damping rate} & 100 $\text{Hz}$ &  \cite{breakspear2006unifying, robinson2002dynamics, marten2009onset}\\ \hline
$\tau$ & \multicolumn{2}{l|}{Time delay due to slow synaptic kinetics of GABA$_B$ }& 50 ${\text{ms}}$ & \cite{marten2009onset}\\ \hline
$\alpha$ & \multicolumn{2}{l|}{Synaptodendritic decay time constant} & 50 ${\text{s}}^{-1}$ & \cite{breakspear2006unifying,marten2009onset,robinson2002dynamics, HBM.HBM20032}\\ \hline
$\beta$ & \multicolumn{2}{l|}{Synaptodendritic rise time constant} & 200 ${\text{s}}^{-1}$ &\cite{breakspear2006unifying,marten2009onset,robinson2002dynamics, HBM.HBM20032}\\ \hline
$\sigma$ & \multicolumn{2}{l|}{Threshold variability of firing rate} & $6$ ${\text{mV}}$ & \cite{breakspear2006unifying,robinson2002dynamics, HBM.HBM20032}\\ \hline
$\phi_{n}$ & \multicolumn{2}{l|}{Nonspecific subthalamic input onto SRN} & $2$ ${\text{mV s}}$ & \cite{breakspear2006unifying,robinson2002dynamics, HBM.HBM20032}\\ \hline
\end{tabular}
\begin{flushleft}
\caption{Model parameters employed in the present study. Unless otherwise noted, we use these default parameter values for simulations.}
\end{flushleft}
\label{tab:label}
 \end{table*}

\subsection*{Data analysis}
In the present study, several data analysis methods are employed to quantitatively evaluate the dynamical states as well as the properties of SWDs generated by the model. To reveal critical transitions between different dynamical states, we perform the bifurcation analysis for several key parameters of the model. For one specific parameter, the bifurcation diagram is simply obtained by plotting the ``stable'' local minimum and maximum values of cortical excitatory axonal fields (i.e., $\phi_e$) over changes in this parameter \cite{breakspear2006unifying, Marc2011}. To this end, all simulations are executed for sufficiently long time (10 seconds of simulation time, after the system reaches its stable time series), and only the local minimum and maximum values obtained from the latter stable time series are used. Using the above bifurcation analysis, we can also easily distinguish different dynamical states for combined parameters. Such analysis technique allows us to further identify different dynamical state regions in the two-parameter space (for example, see Fig.~2D). On the other hand, the power spectral analysis is used to estimate the dominant frequency of neural oscillations. To do this, the power spectral density is obtained from the time series $\phi_e$ (over a period of 10 seconds) by using the fast Fourier transform. Then, the maximum peak frequency is defined as the dominant frequency of neural oscillations. It should be noted that, by combining the results of both the state and frequency analysis, we can outline the SWD oscillation region that falls into the 2-4 Hz frequency range in the two-parameter space (for example, see the asterisk region in Fig.~2E). Moreover, we calculate the mean firing rates (MFRs) for several key neural populations in some figures. To compute the MFRs, all corresponding simulations are performed up to 25 seconds and the data from 5 to 25 seconds are used for statistical analysis. To obtain convincing results, we carry out 20 independent simulations with different random seeds for each experimental setting, and report the averaged result as the final result. Finally, in some cases, we also compute the low and high triggering mean firing rates (TMFRs) for SNr neurons. In the following simulations, we find that the mean firing rate of SNr neurons is increased with the growth of the excitatory coupling strength $v_{p_1\zeta}$, which serves as a control parameter to modulate the activation level of SNr in our work (see the Results section). Based on this property, the low and high TMFRs can be determined by the mean firing rates of SNr neurons occurring at the boundaries of the typical region of 2-4~Hz SWDs (for example, see the black dashed lines in Fig.~3B).

\subsection*{Numerical simulation}
All network simulations are written and performed under the MATLAB environment. The aforementioned dynamical equations are integrated by using the standard fourth-order Runge-Kutta method, with a fixed temporal resolution of $h=0.05$ ms \cite{Butcher2003}. In additional simulations, it turns out that the chosen integration step is sufficiently small to ensure the numerical accuracy of our developed BGCT model. The basic computer code used in the present study is provided as supplementary information to this paper (we also provide a XPPAUT code for comparison \cite{Ermentrout2002}; see Text~S1 and S2).

\section*{Results}
\subsection*{Slow kinetics of ${\text{GABA}}_{\text{B}}$ receptors in TRN create absence seizure activities for the BGCT model}

Previous studies have suggested that the slow kinetics of ${\text{GABA}}_{\text{B}}$ receptors in TRN are a candidate pathological factor contributing to the generation of absence seizures both in animal experiments and biophysical models of corticothalamic network \cite{destexhe1998spike,Hosford92, marten2009onset, Destexhebook2008}. To explore whether this mechanism also applies to the developed BGCT model, we perform one-dimensional bifurcation analysis for the inhibitory coupling strength $-v_{sr}$ and the delay parameter $\tau$, respectively. The corresponding bifurcation diagrams and typical time series of $\phi_e$ are depicted in Figs.~2A-2C, which reveal that different dynamical sates emerge in our system for different values of $-v_{sr}$ and $\tau$. When the coupling strength $-v_{sr}$ is too weak, the inhibition from TRN cannot effectively suppress the firing of SRN. In this case, due to the strong excitation from pyramidal neurons, the firing of SRN rapidly reaches a high level after the beginning of the simulation. Such high activation level of SRN in turn drives the firing of cortical neurons to their saturation states within one or two oscillation periods (region I). As the coupling strength $-v_{sr}$ grows, the inhibition from TRN starts to affect the firing of SRN. For sufficiently long $\tau$, this causes our model to successively undergo two different oscillation patterns. The first one is the SWD oscillation pattern, in which multiple pairs of maximum and minimum values are found within each periodic complex (region II). Note that this oscillation pattern has been extensively observed on the EEG recordings of real patients during absence seizures \cite{panayiotopoulos1997absence}. The other one is the simple oscillation pattern, in which only one pair of maximum and minimum values appears within each periodic complex (region III). However, if the coupling strength $-v_{sr}$ is too strong, the firing of SRN is almost completely inhibited by TRN. In this situation, the model is kicked into the low firing region and no oscillation behavior can be observed anymore (region IV). Additionally, we also find that the model dynamics are significantly influenced by the ${\text{GABA}}_{\text{B}}$ delay, and only sufficiently long $\tau$ can ensure the generation of SWDs in the developed model (see Fig.~2B).

\begin{figure*}[tp]
\centering
\includegraphics[width=15cm]{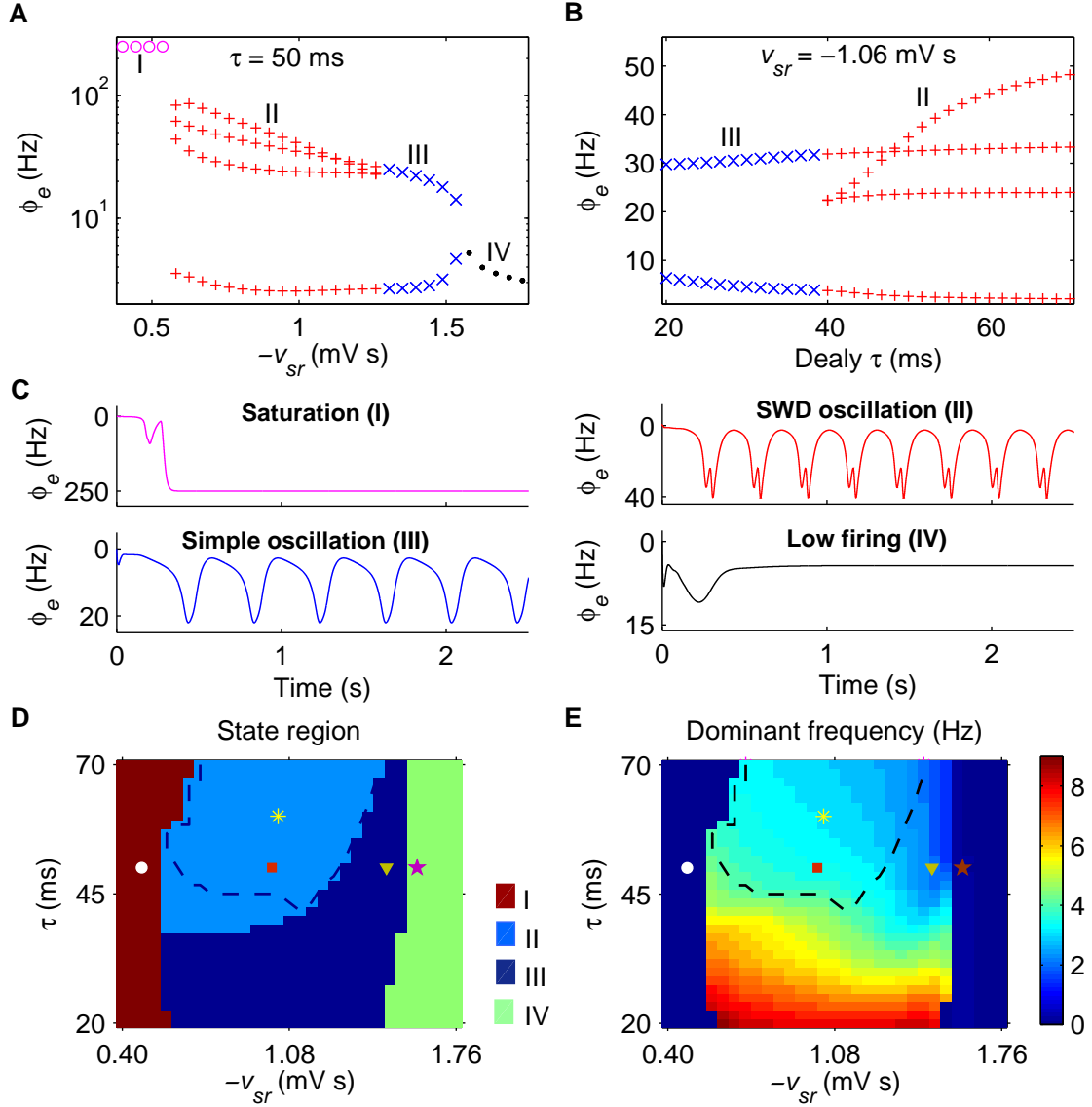}
\caption{\textbf{Absence seizure activities induced by the slow kinetics of ${\text{GABA}}_{\text{B}}$ receptors in TRN.} A, B: Bifurcation diagrams of $\phi_{e}$ as a function of the TRN-SRN inhibitory coupling strength $-v_{sr}$ (A) and the delay parameter $\tau$ (B), respectively. Four different dynamical states can be observed from the time series of $\phi_{e}$, which are: the saturation state (I), the SWD oscillation state (II), the simple oscillation state (III) and the low firing state (IV). C: Typical time series of $\phi_{e}$ correspond to the above four dynamical states. Here we set $\tau=50$ ms and chose $v_{sr}=-0.48$ mV~s (I),  $v_{sr}=-1$ mV~s (II),  $v_{sr}=-1.48$ mV~s (III),  $v_{sr}=-1.6$ mV~s (IV), respectively. The colors in bifurcation diagrams (A) and (B) correspond to the typical time series plotted in (C). D, E: The state analysis (D) and frequency analysis (E) in the ($-v_{sr}, \tau$) panel. Different colors in (D) represent different dynamical state regions, corresponding to those dynamical states given in (A), (B) and (C). The asterisk (``$\ast$'') regions surrounded by black dashed lines in (D) and (E) represent the SWD oscillation regions falling into the 2-4 Hz frequency range. The other symbols in (D) and (E) are linked to parameter values used for different typical time series in (C): I (``$\bullet$''), II (``$\blacksquare$''), III (``$\blacktriangledown$''), and IV (``$\bigstar$''). For all simulations, we set $v_{{p_1}{\zeta}}=0.3$ mV~s.}
\label{fig:Fig.2}
\end{figure*}

To check whether our results can be generalized within a certain range of parameters, we further carry out the two-dimensional state analysis in the ($-v_{sr}, \tau$) panel. As shown in Fig.~2D, the whole ($-v_{sr}, \tau$) panel is divided into four state regions, corresponding to those regions identified above. Unsurprisingly, we find that the BGCT model can generate the SWD oscillation pattern only for appropriately intermediate $-v_{sr}$ and sufficiently long $\tau$. This observation is in consistent with our above finding, demonstrating the generalizability of our above results. To estimate the frequency characteristics of different oscillation patterns, we compute the dominant frequency based on the spectral analysis in the ($-v_{sr}, \tau$) panel. For both the simple and SWD oscillation patterns, the dominant frequency is influenced by $\tau$ and $-v_{sr}$, and increasing their values can both reduce the dominant frequency of neural oscillations (Fig.~2E). However, compared to $-v_{sr}$, our results indicate that the ${\text{GABA}}_{\text{B}}$ delay may have a more significant effect on the dominant oscillation frequency (Fig.~2E). By combining the results in Figs.~2D and 2E, we roughly outline the SWD oscillation region that falls into the 2-4 Hz frequency range (asterisk region). It is found that most of, but not all, the SWD oscillation region is contained in this specific region. Here we emphasize the importance of this specific region, because the SWDs within this typical frequency range is commonly observed during the paroxysm of absence epilepsy in human patients \cite{panayiotopoulos1997absence, crunelli2002childhood}.

Why can the slow kinetics of ${\text{GABA}}_{\text{B}}$ receptors in TRN induce absence seizure activities? Anatomically, the SRN neurons receive the TRN signals from the inhibitory pathway mediated by both ${\text{GABA}}_{\text{A}}$ and ${\text{GABA}}_{\text{B}}$ receptors. Under suitable condition, the double suppression caused by these two types of GABA receptors occurring at different time instants may provide an effective mechanism to create multiple firing peaks for the SRN neurons (see below). Such firing pattern of SRN in turn impacts the dynamics of cortical neurons, thus leading to the generation of SWDs. It should be noted that, during the above processes, both $\tau$ and $-v_{sr}$ play critical roles. In each oscillation period, after the ${\text{GABA}}_{\text{A}}$-induced inhibition starts to suppress the firing of SRN neurons, these neurons need a certain recovery time to restore their mean firing rate to the rising state. Theoretically, if this recovery time is shorter than the ${\text{GABA}}_{\text{B}}$ delay, another firing peak can be introduced to SRN neurons due to the latter ${\text{GABA}}_{\text{B}}$-induced inhibition. The above analysis implies that our model requires a sufficient long ${\text{GABA}}_{\text{B}}$ delay to ensure the occurrence of SWDs. However, as described above, too long $\tau$ is also a potential factor which may push the dominant frequency of SWDs beyond the typical frequency range. For a stronger $-v_{sr}$, the inhibition caused by ${\text{GABA}}_{\text{A}}$ is also strong. In this situation, it is obvious that the SRN neurons need a longer time to restore their firing rate. As a consequent, a relatively longer $\tau$ is required for the BGCT model to ensure the occurrence of SWDs for stronger $-v_{sr}$ (see Fig.~2D).

These findings provide consistent evidence that our developed BGCT model can replicate the typical absence seizure activities utilizing previously verified pathological mechanism. Because we do not change the normal parameter values for basal ganglia during above studies, our results may also indicate that, even though the basal ganglia operate in the normal state, the abnormal alteration within the corticothalamic system may also trigger the onset of absence epilepsy. Throughout the following studies, we set $\tau=50$ ms for all simulations. For this choice, the delay parameter $\tau$ is within the physiological range and modest, allowing the generation of SWD oscillation pattern while preserving its dominant frequency around 3 Hz in most considered parameter regions. It should be noted that, in additional simulations, we have shown that by slightly tuning the values of several parameters our developed BGCT model is also powerful to reproduce many other typical patterns of time series, such as the alpha and beta rhythms (see Figure~S1), which to a certain extent can be comparable with real physiological EEG signals \cite{robinson2002dynamics,HBM.HBM20032}.

\subsection*{Control of absence seizures by the isolated SNr-TRN pathway}

Using the developed BGCT model, we now investigate the possible roles of basal ganglia in controlling absence seizure activities. Here we mainly concentrate on how the activation level of SNr influence the dynamics generated by the model. This is because, on the one hand, the SNr is one of chief output nucleus of the basal ganglia to thalamus, and on the other hand, its firing activity has been found to be highly associated with the regulation of absence seizures \cite{deransart1998role, deransart2002control}. To this end, the excitatory coupling strength $v_{p_1\zeta}$ is employed to control the activation level of SNr and a three-step strategy is pursued in the present work. In this and next subsections, we assess the individual roles of two different pathways emitted from SNr to thalamus (i.e., the SNr-TRN and SNr-SRN pathways) in the control of absence seizures and discuss their corresponding biophysical mechanisms, respectively. In the final two subsections, we further analyze the combination effects of these two pathways on absence seizure control and extend our results to more general cases.

\begin{figure*}[tp]
\center
\includegraphics[width=15cm]{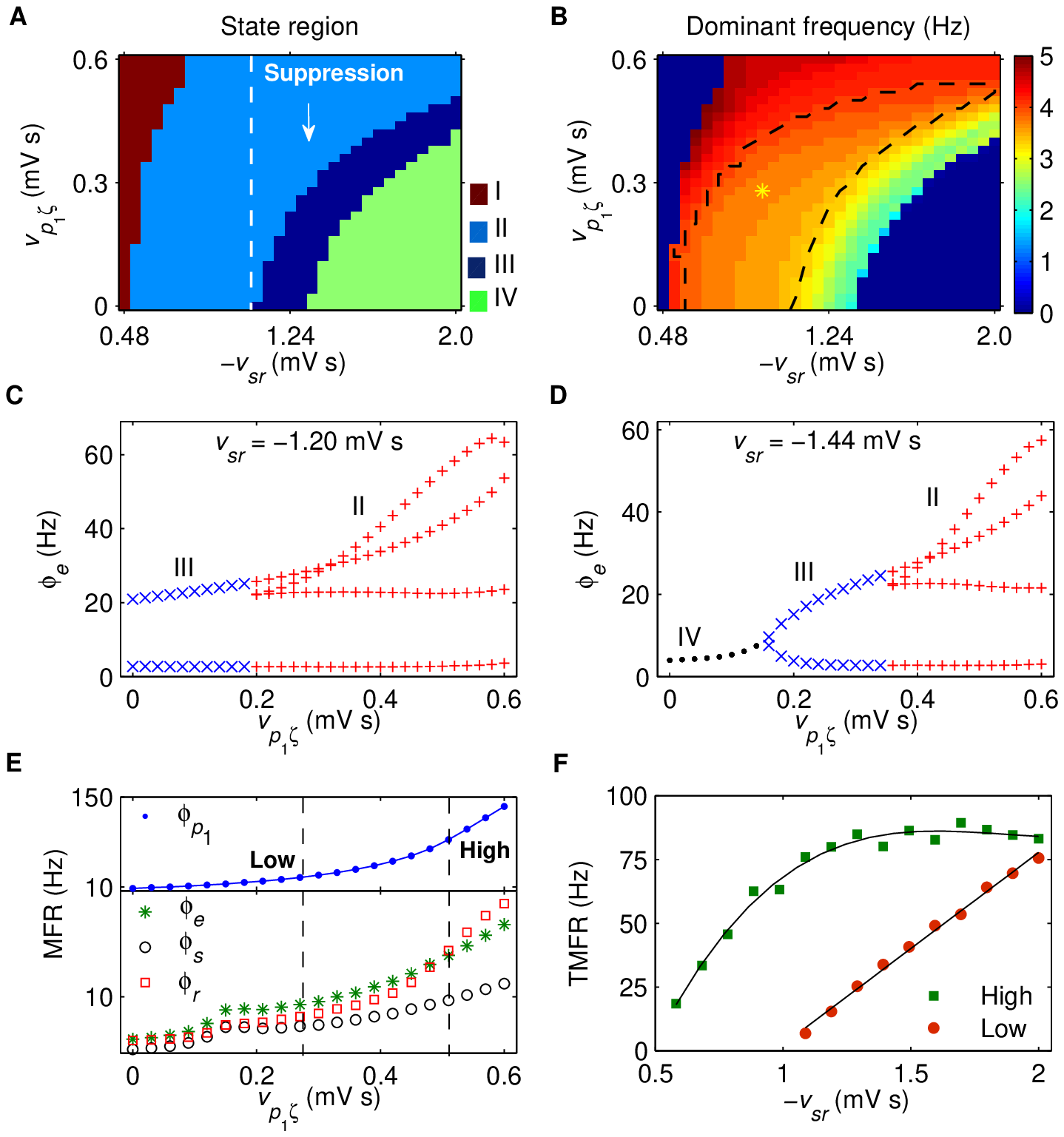}
\caption{\textbf{Control of absence seizures by the isolated SNr-TRN pathway.} A, B: The state analysis (A) and frequency analysis (B) in the ($-v_{sr}, v_{p_1\zeta}$) panel. Here $-v_{sr}$ is the inhibitory coupling strength of the TRN-SRN pathway, whereas $v_{p_1\zeta}$ is the excitatory coupling strength of the STN-SNr pathway. Different colors in (A) represent different dynamical state regions: the saturation region (I), the SWD oscillation region (II), the simple oscillation region (III) and the low firing region (IV). The suppression of SWDs appears to the right of the white dashed line in (A), where the down arrow indicates that the SWD oscillation can be inhibited by decreasing $v_{p_1\zeta}$. The asterisk (``$\ast$'') region surrounded by black dashed lines in (B) denotes the typical 2-4 Hz SWD oscillation region. C, D: Bifurcation diagrams of $\phi_{e}$ as a function of $v_{p_1\zeta}$ for different $-v_{sr}$. The strengths of the inhibitory projections from the TRN to SRN are set as $v_{sr}=-1.20$ mV~s (C) and $v_{sr}=-1.44$ mV~s (D), respectively. Different colors in (C) and (D) represent different dynamical state regions, corresponding to those in phase diagram (A). E: The mean firing rates (MFRs) of several key neural populations as a function of $ v_{p_1\zeta}$, with $v_{sr}=-1.44$ mV~s. Here four neural populations are considered: SNr (``$\cdot$''), excitatory pyramidal neurons (``$\ast$''), SRN (``$\circ$'') and TRN (``$\square$''). Two black dashed lines in (E) represent the occurring positions of the low and high triggering mean firing rates (TMFRs), respectively. F: The low and high TMFRs as a function of $-v_{sr}$. For all simulations, the SNr-SRN pathway is artificially blocked (i.e., $v_{{sp}_1}=0$ mV~s).}
\label{fig:Fig.3}
\end{figure*}

To explore the individual role of the SNr-TRN pathway, we estimate both the state regions and frequency characteristics in the ($-v_{sr}, v_{p_1\zeta}$) panel. Note that during these investigations the SNr-SRN pathway is artificially blocked (i.e., $v_{{sp}_1}=0$ mV~s). With this ``naive'' method, the modulation of absence seizure activities by the SNr-SRN pathway is removed and the effect caused by the SNr-TRN pathway is theoretically amplified to the extreme. Similar to previous results, we find that the whole ($-v_{sr}, v_{p_1\zeta}$) panel can be also divided into four different regions (Fig.~3A). These regions are the same as those defined above. For weak inhibitory coupling strength $-v_{sr}$, increasing the excitatory coupling strength $v_{p_1\zeta}$ moves the model dynamics from the SWD oscillation state to the saturation state. Here we have to notice that the saturation state is a non-physiological brain state even though it does not belong to typical seizure activities. In strong $-v_{sr}$ region, the suppression of SWDs is observed by decreasing the excitatory coupling strength $v_{p_1\zeta}$, suggesting that inactivation of SNr neurons may result in seizure termination through the SNr-TRN pathway (Fig.~3A, right side). For strong enough inhibitory coupling strength $-v_{sr}$, such suppression effect is rather remarkable that sufficiently low activation of SNr can even kick the network dynamics into the low firing region (compare the results in Figs.~3C and 3D).

The SNr-TRN pathway induced SWD suppression is complicated and its biophysical mechanism is presumably due to competition-induced collision. On the one side, the decrease of excitatory coupling strength $v_{p_1\zeta}$ inactivates the SNr (Fig.~3E, top panel), which should potentially enhance the firing of TRN neurons. On the other side, however, increasing the activation level of TRN tends to suppress the firing of SRN, which significantly reduces the firing of cortical neurons and in turn inactivates the TRN neurons. Furthermore, the inactivation of cortical neurons also tends to reduce the firing level of TRN neurons. As the excitatory coupling strength $ v_{p_1\zeta}$ is decreased, the collision caused by such complicated competition and information interactions finally leads to the inactivation for all the TRN, SRN, and cortical neurons (Fig.~3E, bottom panel), which potentially provides an effective mechanism to destabilize the original pathological balance within the corticothalamic system, thus causing the suppression of SWDs.

Indeed, we find that not only the dynamical state but also the oscillation frequency is greatly impacted by the activation level of SNr, through the SNr-TRN pathway. For both the simple and SWD oscillation patterns, increasing the excitatory strength $v_{p_1\zeta}$ can enhance their dominant frequencies. The combined results of Figs.~3A and 3B reveal that, for a fixed $-v_{sr}$, whether the model can generate the SWDs within the typical 2-4 Hz is determined by at least one and often two critical values of $v_{p_1\zeta}$ (Fig.~3B, asterisk region). Because the activation level of SNr is increased with the growth of $ v_{p_1\zeta}$, this finding further indicates that, due to effect of the SNr-TRN pathway, the model might exist the corresponding low and high triggering mean firing rates (TMFRs) for SNr neurons (Fig.~3E, dashed lines). If the long-term mean firing rate of SNr neurons falls into the region between these two TMFRs, the model can highly generate typical 2-4 Hz SWDs as those observed on the EEG recordings of absence epileptic patients. In Fig.~3F, we plot both the low and high TMFRs as a function of the inhibitory coupling strength $-v_{sr}$. With the increasing of $-v_{sr}$, the high TMFR grows rapidly at first and then reaches a plateau region, whereas the low TMFR almost linearly increases during this process. Consequently, it can be seen that these two critical TMFRs approach each other as the inhibitory coupling strength $-v_{sr}$ is increased until they almost reach an identical value (Fig.~3F).

The above findings indicate that the SNr-TRN pathway may play a vital role in controlling the absence seizures and appropriately reducing the activation level of SNr neurons can suppress the typical 2-4 Hz SWDs. The similar antiepileptic effect induced by inactivating the SNr has been widely reported in previous electrophysiological experiments based on both genetic absence epilepsy rats and tottering mice \cite{deransart1998role, deransart2002control, paz2007activity, Kase2012}. Note that, however, in literature such antiepileptic effect by reducing the activation of SNr is presumed to be accomplished through the indirect SNr-TRN pathway relaying at superior colliculus \cite{deransart1998role, deransart2002control}. Our computational results firstly suggest that such antiepileptic process can be also triggered by the direct SNr-TRN GABAergic projections. Combining these results, we postulate that for real absence epileptic patients both of these two pathways might work synergistically and together provide a stable mechanism to terminate the onset of absence epilepsy.

\subsection*{Control of absence seizures by the isolated SNr-SRN pathway}
We next turn on the SNr-SRN pathway and investigate whether this pathway is also effective in the control of absence seizures. Similar to the previous method, we artificially block the SNr-TRN pathway (i.e., $v_{{rp}_1}=0$ mV s) to enlarge the effect of the SNr-SRN pathway to the extreme. Figure~4A shows the two-dimensional state analysis in the ($-v_{sr}, v_{p_1\zeta}$) panel, and again the whole panel is divided into four different state regions. Compared to the results in Fig.~3A, the suppression of SWDs appears in a relatively weaker $-v_{sr}$ region by increasing the excitatory coupling strength $v_{p_1\zeta}$. This finding suggests that the increase in the activation of SNr can also terminate the SWDs, but through the SNr-SRN pathway. For relatively weak $-v_{sr}$ within the suppression region, the SNr-SRN pathway induced suppression is somewhat strong. In this case, the high activation level of SNr directly kicks the network dynamics into the low firing region, without undergoing the simple oscillation state (Fig.~4$\text{C}_2$ and compare with Fig.~4$\text{C}_3$). Note that this type of state transition is a novel one which has not been observed in the SWD suppression caused by the SNr-TRN pathway. For relatively strong $-v_{sr}$ within the suppression region, the double peak characteristic of SWDs generated by our model is weak. In this situation, as the inhibitory coupling strength $-v_{sr}$ is increased, we observe that the network dynamics firstly transit from the SWD oscillation state to the simple oscillation state, and then to the low firing state (Fig.~4$\text{C}_3$).

\begin{figure*}[tp]
\includegraphics[width=15cm]{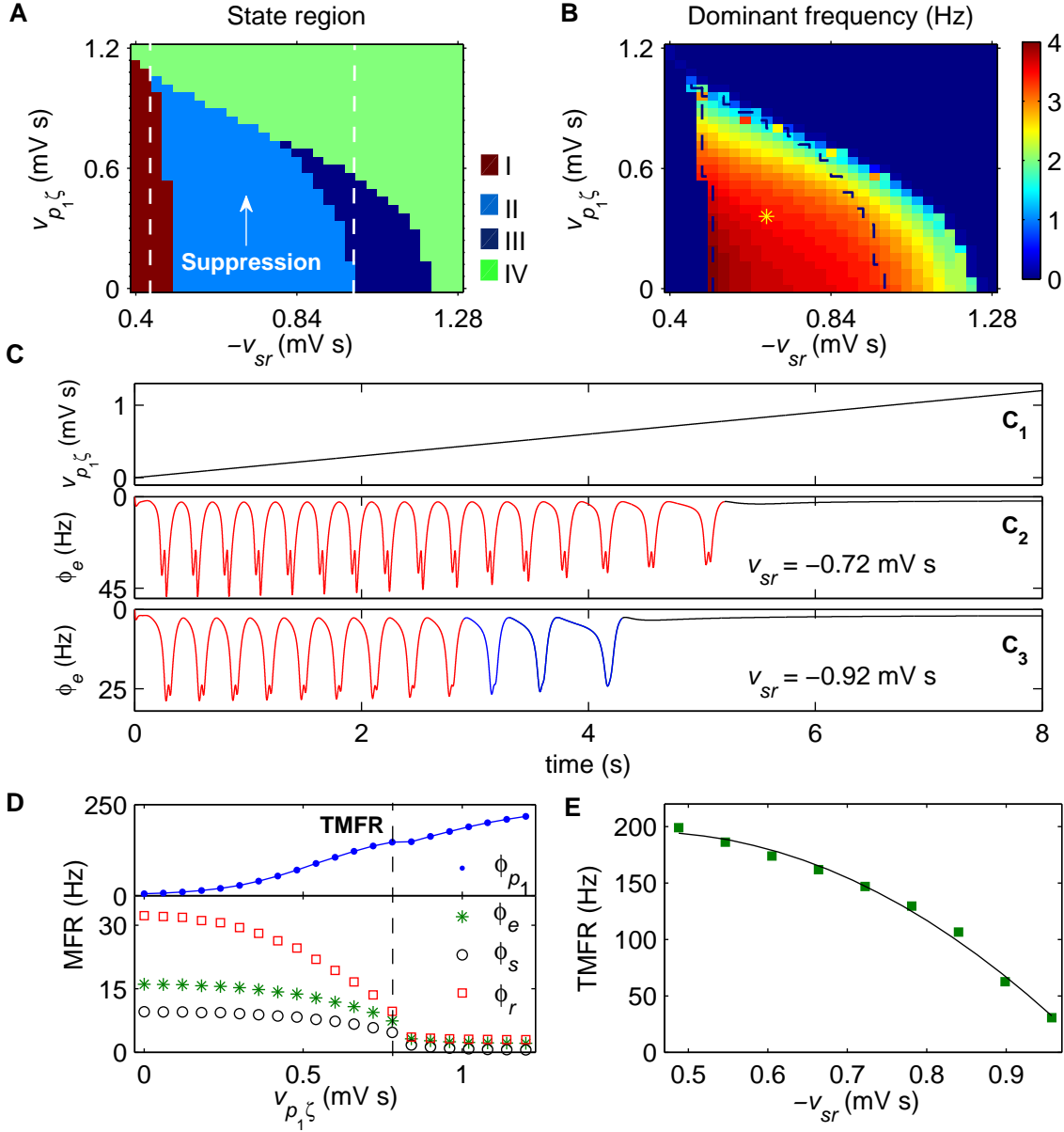}
\caption{\textbf{Control of absence seizures by the isolated SNr-SRN pathway.} A, B: The state analysis (A) and frequency analysis (B) in the ($-v_{sr}, v_{p_1\zeta}$) panel. Here $-v_{sr}$ is the inhibitory coupling strength of the TRN-SRN pathway, whereas $v_{p_1\zeta}$ is the excitatory coupling strength of the STN-SNr pathway. Similar to previous results, four different dynamical state regions are observed: the saturation region (I), the SWD oscillation region (II), the simple oscillation region (III) and the low firing region (IV), which correspond to those defined in Fig.2 (D). The region between two dashed lines in (A) represents the suppression region of SWDs, where the up arrow indicates that the SWD oscillation can be inhibited by increasing $v_{p_1\zeta}$. The asterisk (``$\ast$'') region surrounded by dashed lines in (B) denotes the typical 2-4 Hz SWD oscillation region. C: Two different types of SWD suppressions caused by linearly increasing $v_{p_1\zeta}$. Top (C$_1$): The value of $v_{p_1\zeta}$ as a function of time. Middle (C$_2$): Corresponding $\phi_e$ trace for $v_{sr}=-0.72$ mV~s. Bottom (C$_3$): Corresponding $\phi_e$ trace for $v_{sr}=-0.92$ mV~s. D: The MFRs of several key neural populations as a function of $ v_{p_1\zeta}$, with $v_{sr}=-0.72$ mV~s. Here four neural populations are considered: SNr (``$\cdot$''), excitatory pyramidal neurons (``$\ast$''), SRN (``$\circ$'') and TRN (``$\square$''). The black dashed line in (D) represents the occurring position of TMFR. E: The TMFR as a function of $-v_{sr}$. Note that the SNr-TRN pathway is artificially blocked (i.e., $v_{{rp}_1}=0$ mV~s) for all simulations.}
\label{fig:Fig.4}
\end{figure*}

To understand how the SNr-SRN pathway induced SWD suppression arises, we present the mean firing rates of several key neural populations within the corticothalamic system, as shown in Fig.~4D. It can be seen that increasing the strength $v_{p_1\zeta}$ significantly improves the activation level of SNr (Fig.~4D, top panel), which in turn reduces the firing of SRN neurons (Fig.~4D, bottom panel). The inactivation of SRN neurons further suppresses the mean firing rates for both cortical and TRN neurons (Fig.~4D, bottom panel). These chain reactions lead to the overall inhibition of firing activities in the corticothalamic system, which weakens the double peak shaping effect due to the slow kinetics of ${\text{GABA}}_{\text{B}}$ receptors in TRN. For strong $v_{p_1\zeta}$, such weakening effect is considerable, thus causing the suppression of SWDs. Our results provide the computational evidence that high activation of SNr can also effectively terminate absence seizure activities by the strong inhibition effect from the SNr-SRN pathway. Compared to the SWD suppression induced by the SNr-TRN pathway, it is obvious that the corresponding biophysical mechanism caused by the SNr-SRN pathway is simpler and more direct.

Moreover, our two-dimensional frequency analysis indicates that the dominant frequency of neural oscillations depends on the excitatory coupling strength $v_{p_1\zeta}$ (see Fig.~4B). For a constant $-v_{sr}$, progressive increase of $v_{p_1\zeta}$ reduces the dominant frequency, but not in a very significant fashion. Thus, we find that almost all the SWD oscillation region identified in Fig.~4A falls into the typical 2-4 Hz frequency range (Fig.~4B, asterisk region). Unlike the corresponding results presented in previous subsection, the combination results of Figs.~4A and 4B demonstrate that the BGCT model modulated by the isolated SNr-SRN pathway only exhibits one TMFR for SNr neurons. For a suitably fixed strength $-v_{sr}$, the generation of SWDs can be highly triggered when the mean firing rate of SNr neurons is lower than this critical firing rate (Fig.~4D, dashed line). With the increasing of $-v_{sr}$, we observe that this TMFR rapidly reduces from a high value to a low value (Fig.~4E). Note that this decreasing tendency is in contrast with our previous finding based on the model modulated by the isolated SNr-TRN pathway (compared to the results in Fig.~3F).

Taken together, these observations suggest that increasing the activation level of SNr neurons also significantly influences the dynamics of the corticothalamic system and causes the suppression of absence seizure activities. To the best of our knowledge, this is a new finding that underscores the importance of the direct inhibitory SNr-SRN pathway in controlling and modulating absence seizure activities. It is reasonable to believe that several other external factors, which are able to enhance the activation level of SNr, may also result in the termination of absence seizures due to the similar mechanism.

\subsection*{Competition-induced bidirectional control of absence seizures by the basal ganglia}

\begin{figure*}[tp]
\center
\includegraphics[width=15cm,height=11.8cm]{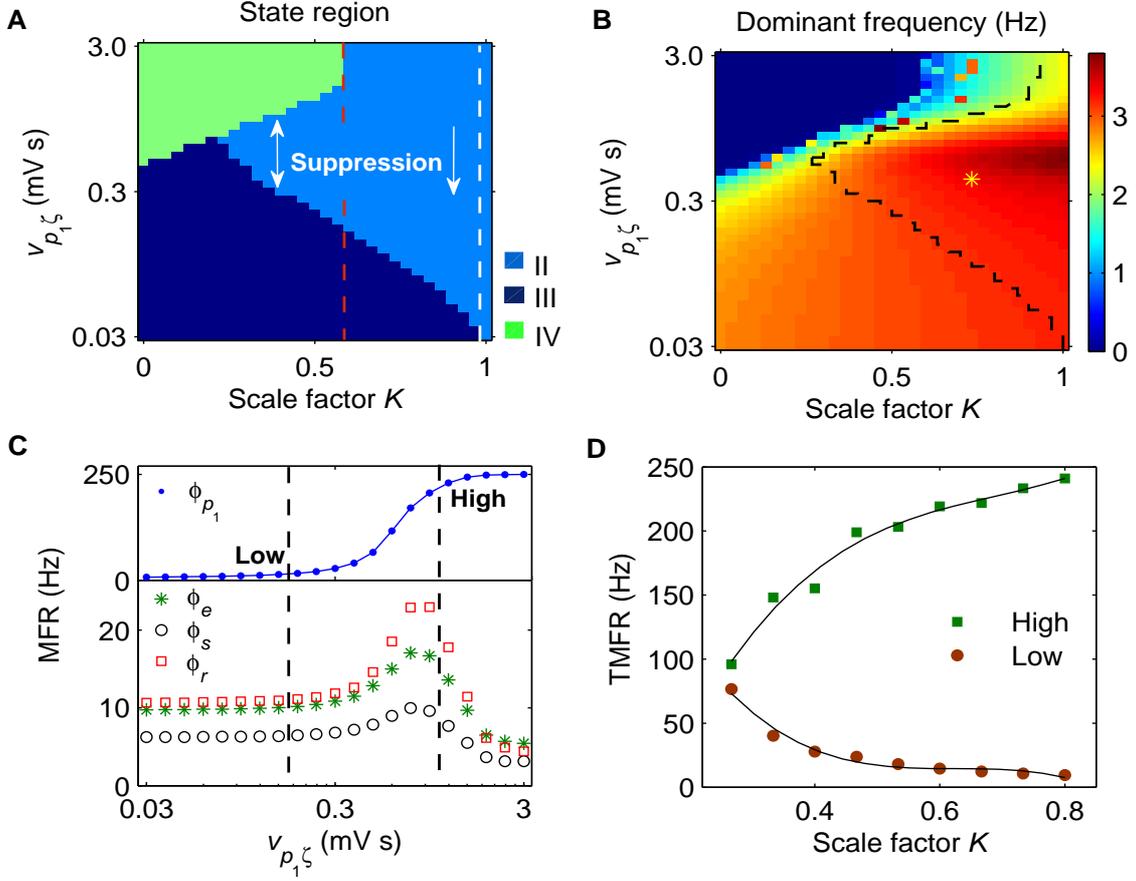}
\caption{\textbf{Control of absence seizures by the combination effects of the SNr-TRN and SNr-SRN pathways.} A, B: The state analysis (A) and frequency analysis (B) in the ($K, v_{p_1\zeta}$) panel. Here $K$ is the scale factor, and $v_{p_1\zeta}$ is the excitatory coupling strength of the STN-SNr pathway. Unlike previous results, only three dynamical state regions are observed in the phase diagram (A): the SWD oscillation region (II), the simple oscillation region (III) and the low firing region (IV), corresponding to the dynamical states defined in Fig.~2(D). For relatively weaker scale factor $K$, both increase and decrease in the activation level of SNr can inhibit the SWDs (double arrow, bidirectional suppression); whereas for sufficiently strong $K$, only reducing the activation level of SNr suppresses the SWDs (single arrow, unidirectional suppression ). In (A), the white dashed line represents the boundary of suppression region, and the red dashed line stands for the demarcation between the bidirectional and unidirectional suppression regions. The asterisk (``$\ast$'') region surrounded by dashed lines in (B) denotes the SWD oscillation region that falls into the 2-4 Hz frequency range. C: The MFRs of several key neural populations as a function of $ v_{p_1\zeta}$, with the scale factor $K=0.6$. Here four neural populations are considered: SNr (``$\cdot$''), excitatory pyramidal neurons (``$\ast$''), SRN (``$\circ$'') and TRN (``$\square$''). Two black dashed lines in (C) represent the occurring positions of low and high triggering mean firing rates (TMFRs), respectively. D: The low and high TMFRs as a function of $K$. For all simulations, the coupling strength of the TRN-SRN pathway is set as $v_{sr}=-1.08$ mV~s.}
\label{fig:Fig.5}
\end{figure*}

\begin{figure*}[tp]
\center
\includegraphics[width=16cm]{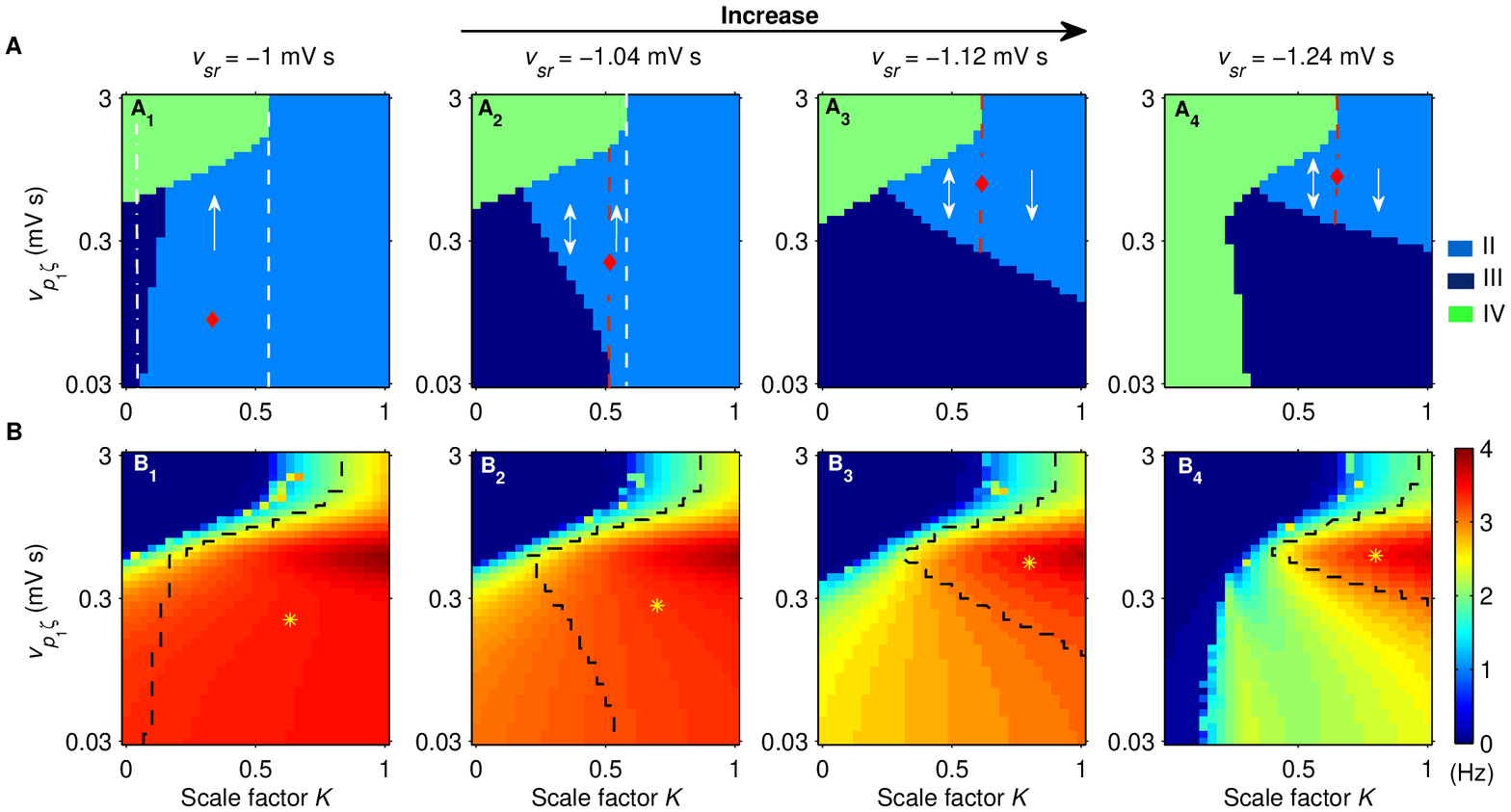}
\caption{\textbf{Effect of inhibitory coupling strength $-v_{sr}$ on the control of absence seizures by the SNr-TRN and SNr-SRN pathways.} Tow-dimensional state analysis (A) and corresponding frequency analysis (B) in the ($K, v_{p_1\zeta}$) panel for different values of $v_{sr}$. Here only three dynamical state regions are observed in (A): the SWD oscillation region (II), the simple oscillation region (III) and the low firing region (IV). In (A$_1$)-(A$_4$), the regions marked by diamonds denote the whole suppression regions of SWDs, the white dashed lines represent the boundaries of suppression regions, and the red dashed lines stand for the demarcations between the bidirectional (double arrow) and unidirectional (single arrow) suppression regions. In (B$_1$)-(B$_4$), the asterisk (``$\ast$'') regions surrounded by dashed lines denote the typical 2-4 Hz SWD oscillation regions. From left to right, the strengths of inhibitory projections from the TRN to SRN are: $v_{sr}=-1$ mV~s (A$_1$, B$_1$), $v_{sr}=-1.04$ mV~s (A$_2$, B$_2$), $v_{sr}=-1.12$ mV~s (A$_3$, B$_3$), and $v_{sr}=-1.24$ mV~s (A$_4$,  B$_4$), respectively. For better showing and comparing the bidirectional suppression regions among different subfigures, here we mainly consider the scale factor for all values of $v_{sr}$ within the same and small interval from 0 to 1. As an additional comparison, both the state analysis and frequency analysis for $v_{sr}=-1.20$ mV~s in a relatively larger $K$ interval is given in Figure~S2.}
\label{fig:Fig.6}
\end{figure*}

So far, we have confirmed that the absence seizure activities generated by the BGCT model can be inhibited by either the isolated SNr-TRN pathway or the isolated SNr-SRN pathway, through different biophysical mechanisms. In real brain, however, both of these pathways should be available and work together at the same time. Thus, an important and naturally arising question is whether these two types of seizure control can coexist in the same network, and if possible, whether this feature can be maintained in certain range of parameters. To address this issue, we introduce a scale factor $K$ and set the coupling strength $v_{rp_1}=Kv_{sp_1}$. By utilizing this method, we can flexibly control the relative coupling strength between the SNr-TRN and SNr-SRN pathways and discuss the combination roles of these two inhibitory pathways in detail.

Figure~5A shows the state analysis in the ($K, v_{p_1\zeta}$) panel with $v_{sr}=-1.08$~mV~s. Unlike the previous results, here we only discover three dynamical state regions, which correspond to: the SWD oscillation state (II), the simple oscillation state (III), and the low firing state (IV). The disappearance of the saturation state is at least due to the following two reasons: (1) the double suppression from the SNr-TRN and SNr-SRN pathways and (2) the relatively strong inhibitory effect from TRN to SRN. As shown in Fig.~5A, the phenomenon of the SWD suppression appears in the strong $K$ region. For relatively weak $K$ within this suppression region, both increasing and decreasing the activation of SNr neurons from the normal level effectively inhibit the generation of SWDs (see Fig.~5A). Such bidirectional suppression behavior can be attributed to the effective competition between the SNr-TRN and SNr-SRN pathways. As the scale factor $K$ is increased, the enhancement of the SNr-TRN pathway breaks the original competition balance between these two inhibitory pathways. During this process, the inhibition from the SNr-TRN pathway progressively dominates the model dynamics. Accordingly, for sufficiently strong $K$, the suppression of SWDs is only found by lowing the activation of SNr (Fig.~5A), which is consistent with our previous critical observation given in Fig.~3A.

However, as described above, the increase in the excitatory coupling strength $v_{p_1\zeta}$ significantly reduces the dominant frequency of SWDs, through the SNr-TRN pathway. Therefore, although enhancing the activation level of SNr neurons cannot inhibit the SWDs directly, it tends to push the dominant frequency of SWDs below 2 Hz (Fig.~5B). By combining the results of Figs.~5A and 5B, we successfully outline the SWD oscillation region that falls into the 2-4 Hz frequency range (asterisk region). In this typical SWD region, the model exhibits both the low and high TMFRs of the SNr neurons for a fixed $K$ (Figs.~5B and 5C). The generation of SWDs within the typical frequency range can be highly triggered if the mean firing rate of SNr neurons is between these two critical TMFRs (Fig.~5C). Nevertheless, due to the competition between the SNr-TRN and SNr-SRN pathways, the firing activities of the TRN, SRN, and cortical neurons become more complicated, compared to the cases induced by any isolated pathway. For a constant $K$, such competition creates a bell-shaped MFR curve for each key neural population by tuning the excitatory coupling strength $v_{p_1\zeta}$, (see Fig.~5C). To further investigate how the scale factor impacts the low and high TMFRs, we plot these two TMFRs as a function of $K$ in Fig.~5D. With the increasing of relative strength $K$, both of these two critical TMFRs are rapidly changed but in different fashions. The high TMFR is increased from a relatively low value to saturation, whereas the low TMFR is reduced from a relatively high value to 0 (see Fig.~5D). Obviously, such opposite tendencies of these two TMFRs are attributed to the combination effects of the SNr-TRN and SNr-SRN pathways.

Furthermore, we also find that both the suppression of SWDs and the typical 2-4 Hz SWD region are shaped by the strength of inhibitory projections from the TRN to SRN. In Figs.~6A and 6B, we perform a series of two-dimensional state and frequency analysis in the ($K, v_{p_1\zeta}$) panel for different values of $-v_{sr}$. When the inhibitory coupling strength $-v_{sr}$ is too weak, the SWDs generated by the corticothalamic system is mainly controlled by the SNr-SRN pathway. In this case, the suppression of SWDs is observed in the intermediate $K$ region and only increasing the activation level of SNr can effectively inhibit the SWDs (Fig.~6A). As the coupling strength $-v_{sr}$ is increased, the inhibition from SNr-TRN pathway starts to influence the model dynamics. This introduces the competition between the SNr-SRN and SNr-TRN pathways, leading to the emergence of the bidirectional suppression of SWDs for intermediate $K$ (Fig.~6A). It is obvious that the higher the coupling strength $-v_{sr}$, the stronger the inhibition effect caused by the SNr-TRN pathway. With increasing the value of $-v_{sr}$, such strengthened inhibition significantly moves the boundary of the low TMFR toward higher values of $v_{p_1\zeta}$, and thus notably shrinks the region of SWDs within the typical frequency range of 2-4 Hz (Fig.~6B).

These above observations emphasize the importance of the combination role of both the SNr-TRN and SNr-SRN pathways on the control of absence seizure activities. Quite remarkably, we observe that the bidirectional suppression of SWDs emerges under suitable conditions, that is, both increasing and reducing the activation levels of SNr neurons effectively suppress the SWDs. Such bidirectional suppression is determined and modulated by both the relative strength of these two inhibitory pathways and the strength of inhibitory projections from the TRN to SRN. This novel finding indicates the possible bidirectional control of absence seizures by the basal ganglia, which is induced by the competition between the SNr-TRN and SNr-SRN pathways. In additional simulations, we have found that several other factors, which can effectively change the activation level of SNr, can also lead to the bidirectional suppression of SWDs due to the similar mechanism, further demonstrating the generality of our results.

\subsection*{Bidirectional control of absence seizures by the basal ganglia might be extended to other pathological factors}

\begin{figure}[tp]
\center
\includegraphics[width=16cm]{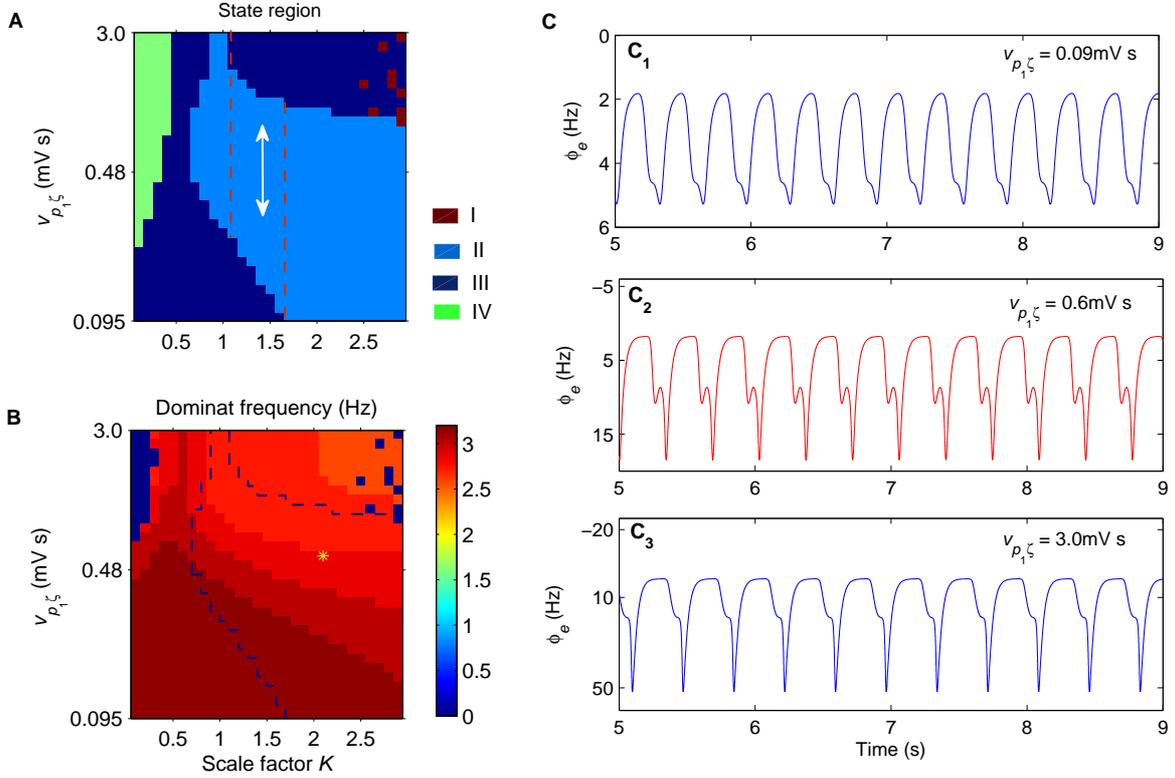}
\caption{\textbf{An example illustrating that the bidirectional control of absence seizures by the basal ganglia is also available for the corticothalamic loop transmission delay induced SWDs.} A, B: Tow-dimensional state analysis (A) and corresponding frequency analysis (B) in the ($K, v_{p_1\zeta}$) panel. Here $K$ is the scale factor, and $v_{p_1\zeta}$ represents the excitatory coupling strength of the STN-SNr pathway. Four different dynamical states are observed in (A): the saturation state (I), the SWD oscillation state (II), the simple oscillation state (III) and the low firing state (IV), which also correspond to previous figures. In (A), the region between two red dashed lines denotes the main bidirectional suppression region of SWDs, where the double arrow represents both increasing and decreasing the excitatory coupling strength $v_{p_1\zeta}$ can inhibit the SWDs. In (B), the asterisk (``$\ast$'') region surrounded by dashed lines denotes the typical 2-4 Hz SWD oscillation region. (C) Three typical time series of $\phi_{e}$ for different values of $v_{p_1\zeta}$, with $K=1.3$. Here we choose $v_{p_1\zeta}=0.09$~mV~s (top), $v_{p_1\zeta}=0.6$~mV~s (middle) and $v_{p_1\zeta}=3.0$~mV~s (bottom), respectively. In all simulations, we set $t_0=80$ ms, $v_{es}=3.2$  mV~s, $v_{se}=3.4$  mV~s, $v_{re}=1.6$ mV~s, $v_{sr}=-1.76$ mV~s and $\phi_{n}=8.0$ mV~s.}
\label{fig:Fig.7}
\end{figure}

In above subsections, we focused on one specific pathological factor and have computationally shown that the basal ganglia may bidirectionally control and modulate the typical absence seizure activities (i.e., the SWDs) induced by the slow synaptic kinetics of ${\text{GABA}}_{\text{B}}$ in TRN. A natural question to ask is whether such bidirectional control feature caused by the basal ganglia is a generalized regulatory mechanism for absence seizures. We argue that this might be true. To check this postulation, at least one additional SWD generation mechanism should be introduced into our model, and we need to examine whether the bidirectional control of absence seizures by the basal ganglia is also available for the new pathological factor. In literature, there are several other theories that are associated with the generation mechanisms of SWDs. A boldly accepted one is related to the transmission delay between the cerebral cortex and thalamus and, specifically, it has been found that suitably choosing such transmission delay can drive the corticothalamic system produce the SWDs \cite{breakspear2006unifying, robinson1997propagation}. To apply this pathological factor in our model, here we block the ${\text{GABA}}_{\text{B}}$ pathway from TRN to SRN, and consider a bidirectional $t_0/2$ transmission delay between the cerebral cortex and thalamus, as that used in previous studies \cite{breakspear2006unifying, robinson1997propagation}. Additionally, several coupling strengths within the corticothalamic loop are also needed to be adapted, because such transmission delay induced SWDs require strong interactions between the cerebral cortex and thalamus \cite{breakspear2006unifying, robinson1997propagation}. The new added and modified model parameters that we used in this subsection are as follows \cite{breakspear2006unifying, robinson1997propagation}: $t_0=80$ ms, $v_{es}=3.2$  mV~s, $v_{se}=3.4$  mV~s, $v_{re}=1.6$ mV~s, $v_{sr}=-1.76$ mV~s and $\phi_{n}=8.0$ mV~s. For simplicity, we term the current model as the modified model in the following studies.

Figures~7A and 7B show an example pair of state analysis and frequency analysis in the ($K, v_{p_1\zeta}$) panel, respectively. As expected, due to the competition between the SNr-TRN and SNr-SRN pathways, we observe the significant bidirectional control feature for intermediate scale factor $K$ (Fig.~7A, the region between dashed lines). In this bidirectional region, both enhancing and lowing the excitatory coupling strength $v_{p_1\zeta}$ push the model dynamics from the SWD oscillation state into the simple oscillation state (Fig.~7C), thus inhibiting the generation of SWDs. This finding supports our above hypothesis that under suitable conditions the basal ganglia may control and modulate the absence seizure activities bidirectionally. However, compared to the results in Fig.5, we find that the bidirectional region appears in a relatively larger $K$ region for the modified model, and increasing the coupling strength $v_{p_1\zeta}$ cannot kick the model dynamics into the low firing state as well. This is not so surprising because these two SWD generation mechanisms that we used are similar but not completely identical. As we introduced above, the SWDs induced by the corticothalamic loop transmission delay require relatively stronger interactions between the cerebral cortex and thalamus, which essentially weaken the inhibitory effect from the SNr neurons. This might lead that the firing of SRN neurons cannot be fully suppressed even when the activation level of SNr reaches its saturation state. Another important finding that we discover here is that the dynamics of the modified model become complicated for large $K$ and strong $v_{p_1\zeta}$ (see Fig.~7A, upper right). In additional simulations, we further perform a series of state analysis for the modified model using the same group of parameter values but different random initial conditions (see Figure~S3). The corresponding results indicate that the modified model shows bistability (the simple oscillation state or the saturation state) in the large $K$ and strong $v_{p_1\zeta}$ region, and final model dynamics significantly depend on the initial conditions.

In conclusion, these findings further stress the combination role of the inhibitory SNr-TRN and SNr-SRN pathways on the control of absence seizure activities. By combining all of our results, we postulate that the bidirectional control by the basal ganglia is possible a generalized regulatory mechanism for absence seizures and may be extendable to other pathological factors, even though the detailed bidirectional control behaviors may not be completely identical for different pathological factors.

\section*{Discussion}
Using a mean-field  macroscopic model that incorporates the basal ganglia, cerebral cortex and thalamus, we presented here the first investigation on how the basal ganglia control the absence epilepsy through the projections directly emitted from the SNr to several key nuclei of thalamus. Through simulations, we demonstrated that the absence seizure activities induced by the slow synaptic kinetics of ${\text{GABA}}_{\text{B}}$ in TRN can be inhibited by either the isolated SNr-TRN pathway or the isolated SNr-SRN pathway via different biophysical mechanisms. More importantly, our results showed that under certain conditions these two types of seizure control can coexist in the same network, suggesting that both decreasing and increasing the activation levels of SNr may considerably suppress the generation of SWDs. Theoretically, such bidirectional control of absence seizures by basal ganglia is due to the effective competition between the SNr-TRN and SNr-SRN pathways, which might be a generalized mechanism for regulating absence seizure activities and can be extended to other pathological factors. In addition, our detailed frequency analysis also indicated that, depending on different system conditions, the developed model may exist low, high or both TMFRs for the SNr neural population for triggering the typical 2-4 Hz SWDs.

These results are at least partly in agreement with former experimental observations. Previously, experimental studies based on electrophysiological recordings have established the linkage that reducing the activation of SNr neurons from the normal level can effectively suppress the SWDs in different rodent animal models \cite{deransart1998role, deransart2002control, paz2007activity, Kase2012}. Such antiepileptic effect was supposed to be attributed to the indirect pathway of the SNr to TRN relaying at superior colliculus. The results presented in this work also demonstrated that decreasing the SNr activity is an effective approach that terminates the SWDs. However, it is important to note that in our model the similar antiepileptic effect is triggered by the direct GABAergic projections from SNr to TRN. Presumably, this is because the SNr has overall inhibitory impacts on TRN via both indirect and direct pathways. In the brain of absence epileptic patients, both of these two pathways might work together and provide a stable and endogenous mechanism to terminate the paroxysm of absence epilepsy.

Our model further makes prediction that increasing the activation of SNr from the normal level may also suppress SWDs. In previous experimental studies, there still lacks sufficient evidence to support this viewpoint. We speculate that this might be because the relative strength of the SNr-TRN and SNr-SRN pathways for rodent animals is generally high or at least not too low. Under this condition, the BGCT system for most rodent animals does not operate in the bidirectional control region, thus the suppression of SWDs caused by activating SNr is difficult to be observed in normal experiments. According to our results, the activation of SNr induced SWD suppression might appear by suitably tuning the relative strength between these two pathways. Further experiments based on animal models will be necessary to validate this prediction and characterize the detailed nature of the SWD suppression induced by the SNr-SRN pathway. Even so, the above prediction from our computational study might provide an alternative approach for terminating absence seizures.

An interesting and important question is: can real brain utilize this bidirectional modulation mechanism to control the paroxysm of absence epilepsy? Our answer is that it is possible, if the real brain has some mechanisms to automatically adjust the balance between the SNr-TRN and SNr-SRN pathways. Theoretically, there are several possible biological mechanisms and one of them is discussed as follows. Experimental data have uncovered that synapses conduct signals in an unreliable fashion, which is due to the probabilistic neurotransmitter release of synaptic vesicles \cite{Raastad1992, Rosenmund1993, Allen1994}.  It has been shown that the transmission failure rate at a given synapse generally tends to exceed the fraction of successful transmission, and in some specific cases it can be even higher than 0.9 \cite{Raastad1992, Rosenmund1993, Allen1994}. Interestingly, recent studies indicated that such synaptic unreliability may play critical functional roles in neural processing and computation \cite{Goldman2002, Guo2011, Guo2012}. For patients with absence seizures, the suitable competition between the SNr-TRN and SNr-SRN pathways can be theoretically achieved by properly tuning the synaptic transmission rate of these pathways. This might be an important underlying mechanism and has significant functional advantages, because it does not require the changes of related anatomical structures and connection densities in the brain. However, we should notice that such tuning is not easy. Specifically, it requires the cooperation among neurons and needs to change the synaptic transmission rates collectively for most of relevant synapses. From the functional perspective, this mechanism may be associated with the self-protection ability of brain. After a long time of evolution, it is reasonable to suppose that our brain might have the abilities to use this type of plasticity-like mechanism to achieve complicated self-protection function during absence seizures~\cite{Guo2012}. Nevertheless, additional well-designed experiments are still needed to test whether our proposed hypothesis is correct.

The current results highlight the functional roles of basal ganglia in the control of absence seizures and might offer physiological implications on different aspects. First, the termination of absence seizures by activating SNr neurons through the SNr-SRN pathway might inspire the treatment of refractory absence seizures. Although the absence epilepsy is one typical benign epilepsy, a considerable proportion of patients yet may fail to achieve freedom from absence seizures and become refractory to multiple antiepileptic drugs \cite{Perry2012,Bouma1996}. For those patients, one possible reason is that the strengths of their direct and indirect SNr-TRN pathways are somewhat weak, compared to the other patients. In this case, reducing the activation level of SNr may lose efficacy in the suppression of SWDs and, contrarily, increasing the firing of SNr neurons may stop the absence seizures. Second, our results might be generalized to the GPi neurons. In addition to the SNr, the GPi is also an important output structure of basal ganglia. The SNr often works in unison with GPi, since they have closely related inputs and outputs, as well as similarities in cytology and function \cite{van2009amean, van2009bmean, BarGad2003}. It is thus reasonable to infer that the activation level of GPi might serve a similar bidirectional role in modulating the absence seizures. Third, the results presented in this study may provide new insight into the deep brain stimulation therapy on absence seizures \cite{Vercueil1998}. In previous studies, it has been demonstrated that absence seizures can be inhibited by suitably applying the deep brain stimulation to STN. Our results not only support the traditional viewpoint, but also indicate that the SNr may be a more direct therapeutic target, compared to STN. We thus infer that the inhibition of typical absence seizure activities can also be accomplished by appropriately stimulating the SNr neurons. Finally, the main goal of this study was to explore the control role of basal ganglia in absence seizures, but not limited to this specific epilepsy. Several other types of epilepsies, such as the juvenile myoclonic epilepsy \cite{Krampfl2005} and the generalized tonic-clonic epilepsy \cite{Ilie2012}, are also highly associated with the corticothalamic system and mediated by GABA receptors. If our above findings on absence seizures capture the real fact, such bidirectional control feature by basal ganglia may be also available for these types of epilepsies.

Although our developed model is powerful enough to suggest some functional roles of basal ganglia in controlling and modulating absence seizures, we have to admit that this biophysical model is simplified and idealized. The limitations of this model and possible extensions in future studies need to be discussed. First, more detailed models based on the spiking neurons are able to introduce much more complexities, such as ionic dynamics, firing variability and connection property, to the system. In previous studies, it has been observed that ion concentrations inside the cell and in the extracellular space as well as the firing dynamics of neurons are changed during epileptic seizures \cite{Frohlich2010, Freyer2011, Volman2011}. It has been also found recently that the connection property of neural systems in different scales (for example, neuronal networks at microscopic scales and brain region networks at macroscopic scales) also contribute to the generation of seizure-like activities \cite{Netoff2004, Ponten2007}. Therefore, we cannot exclude other possible interesting results by using more detailed modelling methodologies. Moreover, due to lack of necessary data, our model does not include the indirect pathway from the SNr to TRN, relaying at the intermediate and superficial layers of the caudal superior colliculus \cite{NailBoucherie2002}. Although we have inferred that both the direct and indirect SNr-TRN pathways play the similar effects on the control of absence seizures, further studies based on detailed anatomical data need to be investigated in the future work. Finally, it should be emphasized that in the present study we mainly focus on the SWDs induced by the slow kinetics of ${\text{GABA}}_{\text{B}}$ receptors in TRN. Even though we also showed that the similar bidirectional control behavior can be observed for another pathological factor (i.e., the corticothalamic loop transmission delay), more detailed analysis on this aspect is still needed. As we know, several other pathological factors, such as the increased T-type Ca$^\text{2+}$ current in thalamocortical neurons \cite{crunelli2002childhood}, may also lead to the generation of SWDs. Nevertheless, it is still unknown and deserves to be further examined whether our proposed bidirectional control by basal ganglia is also available for these other pathological factors.

Theoretically, a better choice for mimicking the slower dynamics of ${\text{GABA}}_{\text{B}}$ currents should use relatively smaller values of $\alpha$ and $\beta$ at the relevant synapses. However, we have to notice that the mean-filed theory used in the present study is under an assumption of uniform $\alpha$ and $\beta$ parameters for all incoming connections of one neural population (see Eq.~3). To our knowledge, introducing different $\alpha$ and $\beta$ to different types of connections will make such simple form of differential operator presented in Eq.~3 become invalid, thus making the dendritic filtering of incoming signals mathematically intractable in the current mean-field framework. Considering this, we employed a delay parameter $\tau$ to represent the order of magnitude of rise and decay time of ${\text{GABA}}_{\text{B}}$ currents in our model, as the method used in a recent neural mean-field modelling study \cite{marten2009onset}. Essentially, more detailed models based on spiking neurons will allow us to consider different dynamical processes for different types of synapses \cite{Guo2012,Wang2011}. We are currently trying to extend our above-identified results to a relevant spiking neural network and the corresponding results will be presented elsewhere.

It should be also pointed out that traditional modelling studies on absence seizures mainly focus on the possible generation mechanism of the typical absence seizure activities, i.e., SWDs, within the corticothalamic system, but little research addresses the regulatory mechanisms of absence seizures by basal ganglia. Inspired by previous experimental observations \cite{deransart1998role, deransart2002control, paz2007activity, Kase2012} and recent modelling studies on Parkinson's disease \cite{van2009amean, van2009bmean}, here we introduced the basal ganglia to the traditional mean-field model of corticothalamic system proposed by Robinson and his collaborators, and firstly explored how the basal ganglia control the SWDs through the computational approach. Theoretically, several other models that describe the macro-scale dynamics about the activity of neural ensembles, such as the Wilson-Cowan model \cite{Wilson72, Wilson73, Taylor2013} and the models based on dynamic causal modelling \cite{Friston03, Marreiros10, Pinotsis12}, can be also used to deal with the similar problem but may need to re-estimate parameters according to different model assumptions.

In summary, we have performed mechanistic studies and investigated the detailed roles of basal ganglia in the control of absence seizures. We have computationally demonstrated that the SWDs generated by the developed model can be terminated and modulated by both decreasing and increasing the activation levels of SNr. Our results provide the first evidence on the bidirectional control of absence seizures by basal ganglia. This finding might deepen our conventional understanding about the functional roles of basal ganglia in the controlling and modulating of absence seizures. For patients with absence epilepsy, these results in turn indicate that the loss of several basal ganglia functions, especially the functions related to the SNr neurons, might further aggravate the attacks of absence epilepsy. We hope that predictions from our systematic model investigation can not only inspire testable hypotheses for future electrophysiological experiments but also provide additional therapeutic strategies for absence seizures.

\section*{Acknowledgements}
We sincerely thank Dr.~Dan Wu and Dr.~Yangsong Zhang for valuable comments on early versions of this manuscript. This work is supposed by the 973 project (No.~2011CB707803), the National Natural Science Foundation of China (No.~61201278, No.~81071222, No.~81371636, No.~81330032, and No.~91232725), and the 111 project (B12027).

\section*{Author Contributions}
Conceived and designed the experiments: MC DG TW YX PX PAVS DY. Performed the experiments: MC DG CL. Analyzed the data: MC DG TW WJ YX DY. Wrote the paper: MC DG YX CL PAVS DY.

\newpage
\section*{Supporting Information}
Figure S1: \textbf{Several other typical time series of $\phi_e$ (left frames) and their corresponding spectra (right frames) generated by our developed BGCT model}. To a certain extent, these model time series are comparable with real physiological EEG signals: eyes-open (A) and (B), alpha rhythm (C), beta rhythm (D), coexistence of alpha and beta rhythms (E), and polyspike and wave (F). Here we adjust the excitatory corticothalamic coupling strength $v_{se}$ and delay parameter $\tau$ for reproducing these time series. The detailed parameter values used in our simulations are: $v_{se}=0.6$~mV~s and $\tau=32$~ms (A), $v_{se}=1.1$~mV~s and $\tau=16$~ms (B), $v_{se}=1.2$~mV~s and $\tau=16$~ms (C), $v_{se}=1.25$~mV~s and $\tau=3$~ms (D), $v_{se}=1.22$~mV~s and $\tau=16$~ms (E), and $v_{se}=2.2$~mV~s and $\tau=70$~ms (F), respectively. In addition, we also introduce a certain level of gaussian white noise into $\phi_{n}$, with the mean $\langle\phi_{n}\rangle$=2.0~mV~s and standard deviation $\sigma(\phi_{n})=0.0141$~mV~s$^{1/2}$. Note that more types of model time series might be observed by further tuning these critical parameter values.

Figure S2: \textbf{An example representation of the bidirectional control of absence seizures by the basal ganglia in a relatively larger scale factor interval.} The state analysis (A) and frequency analysis (B) in the ($K,v_{p_1\zeta}$) panel, with the inhibitory coupling strength $v_{sr}=-1.2$~mV~s. Here $K$ is the scale factor, and $v_{p_1\zeta}$ is the excitatory coupling strength of the STN-SNr pathway used to control the activation level of SNr neurons. Similar to the results in Figs.~5A and 6A, only three dynamical state regions are observed in the phase diagram (A): the SWD oscillation region (II), the simple oscillation region (III) and the low firing region (IV). In (A), the region marked by diamond denotes the whole suppression regions of SWDs, the white dashed line represents the boundary of suppression region, and the red dashed line stands for the demarcation between the bidirectional (double arrow) and unidirectional (single arrow) suppression regions. In (B), the asterisk (``$\ast$'') region surrounded by dashed lines denotes the typical 2-4 Hz SWD oscillation region. Compared to the results shown in Figs.~5 and 6, here we consider a relatively larger scale factor interval from 0 to 2.3.

Figure S3: \textbf{A series of two-dimensional state analysis in the ($K,v_{p_1\zeta}$) panel for the modified model.} In (A)-(H), we use the same group of parameter values but different random initial conditions for simulations. Similar to previous results, four different dynamical states are observed: the saturation state (I), the SWD oscillation state (II), the simple oscillation state (III) and the low firing state (IV). In each subfigure, the region between two red dashed lines denotes the main bidirectional suppression region of SWDs, where the double arrow represents that both increasing and decreasing $v_{p_1\zeta}$ can inhibit the generation of SWDs. The results given in (A)-(H) indicate that the modified model shows bistability (the simple oscillation state or the saturation state) in the large $K$ and strong $v_{p_1\zeta}$ region, and the final dynamics of the modified model significantly depend on the initial conditions. Note that in all simulations, we set $t_0=80$~ms, $v_{es}=3.2$~mV~s, $v_{se}=3.4$~mV~s, $v_{re}$=1.6~mV~s, $v_{sr}=-1.76$~mV~s and $\phi_{n}=8.0$~mV~s.

Text S1: \textbf{Supporting information code.}

Text S2: \textbf{Supporting information code.}
\newpage
\begin{figure}[p]
\center
\includegraphics[width=16cm]{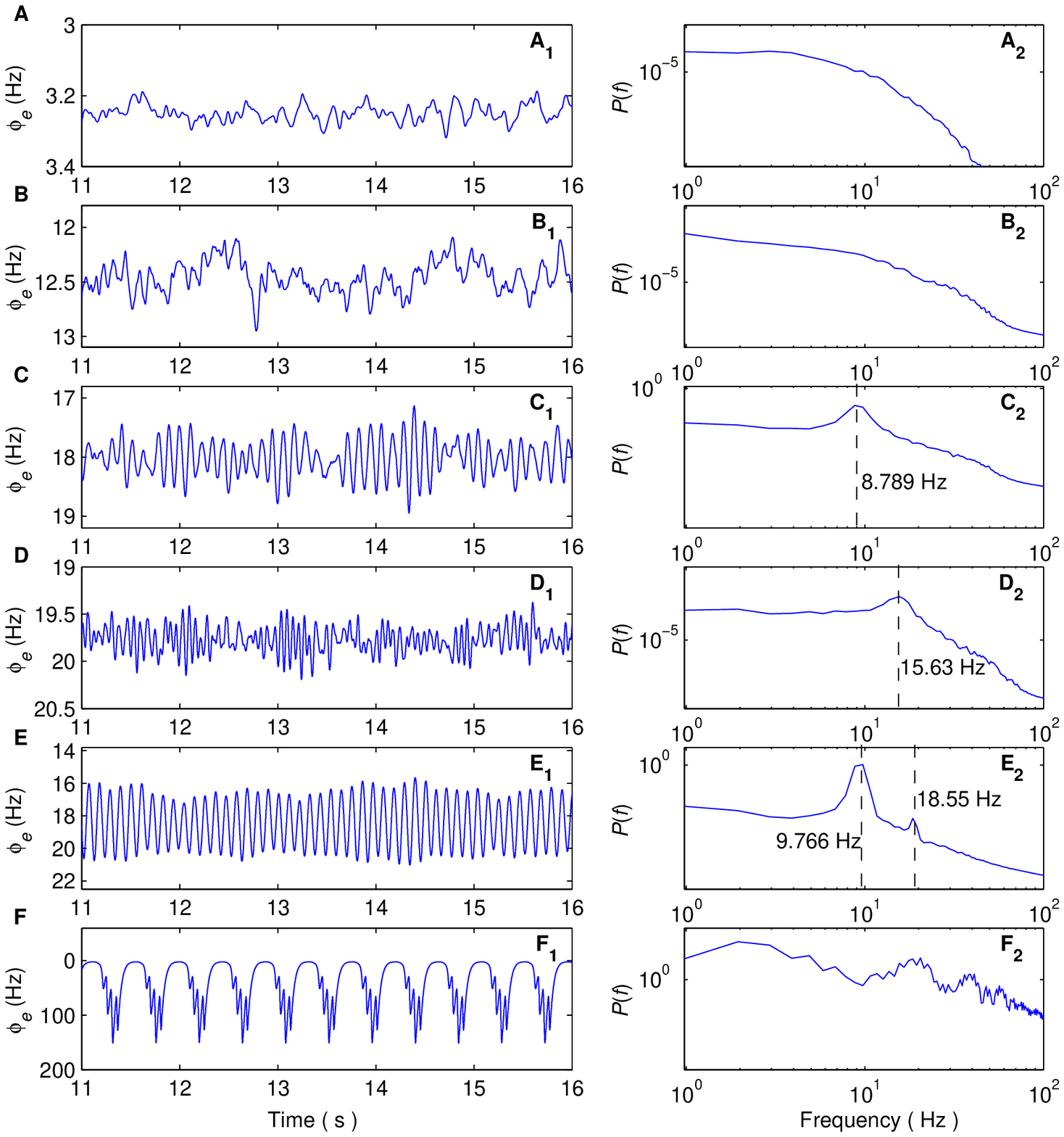}
\textbf{[Figure S1 legend]}
\label{fig:Fig.S1}
\end{figure}

\newpage
\begin{figure}[p]
\center
\includegraphics[width=16cm]{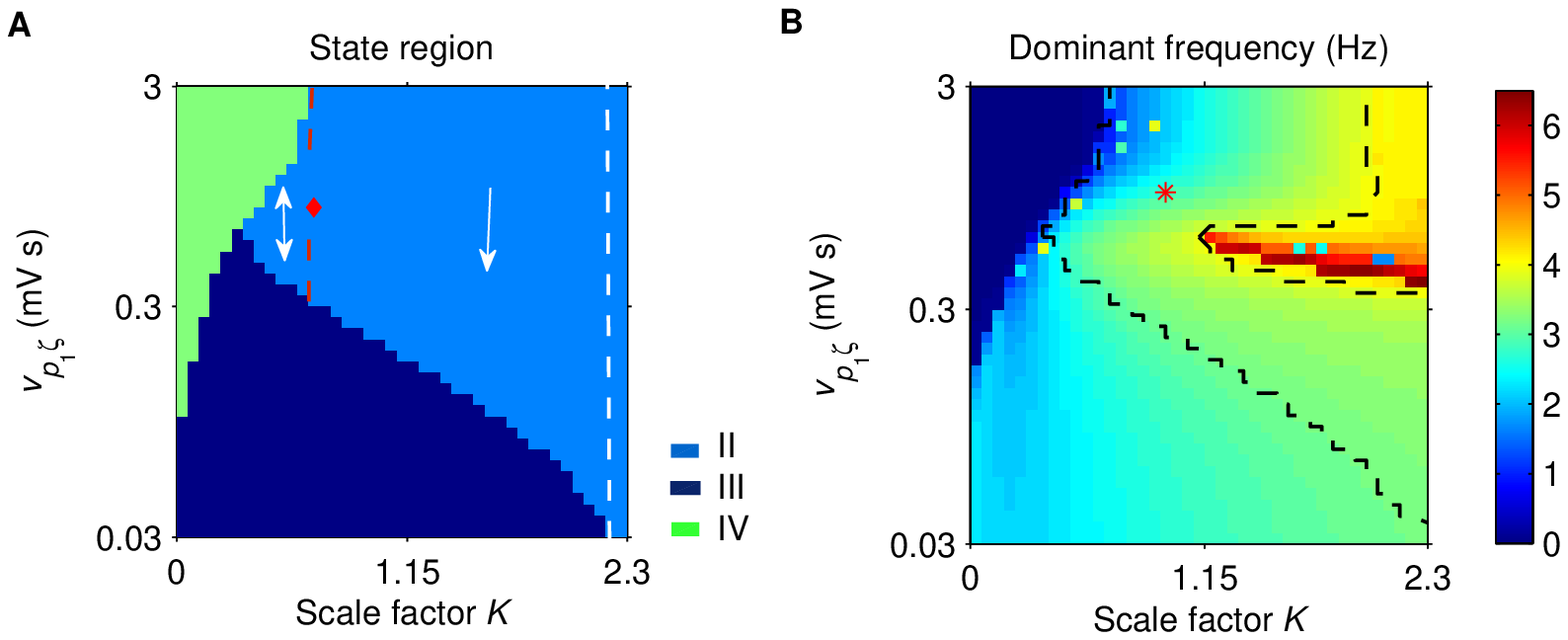}
\textbf{[Figure S2 legend]}
\label{fig:Fig.S2}
\end{figure}

\newpage
\begin{landscape}
\begin{figure}[p]
\renewcommand{\baselinestretch}{1.5}
\center
\includegraphics[width=21cm,height=11cm]{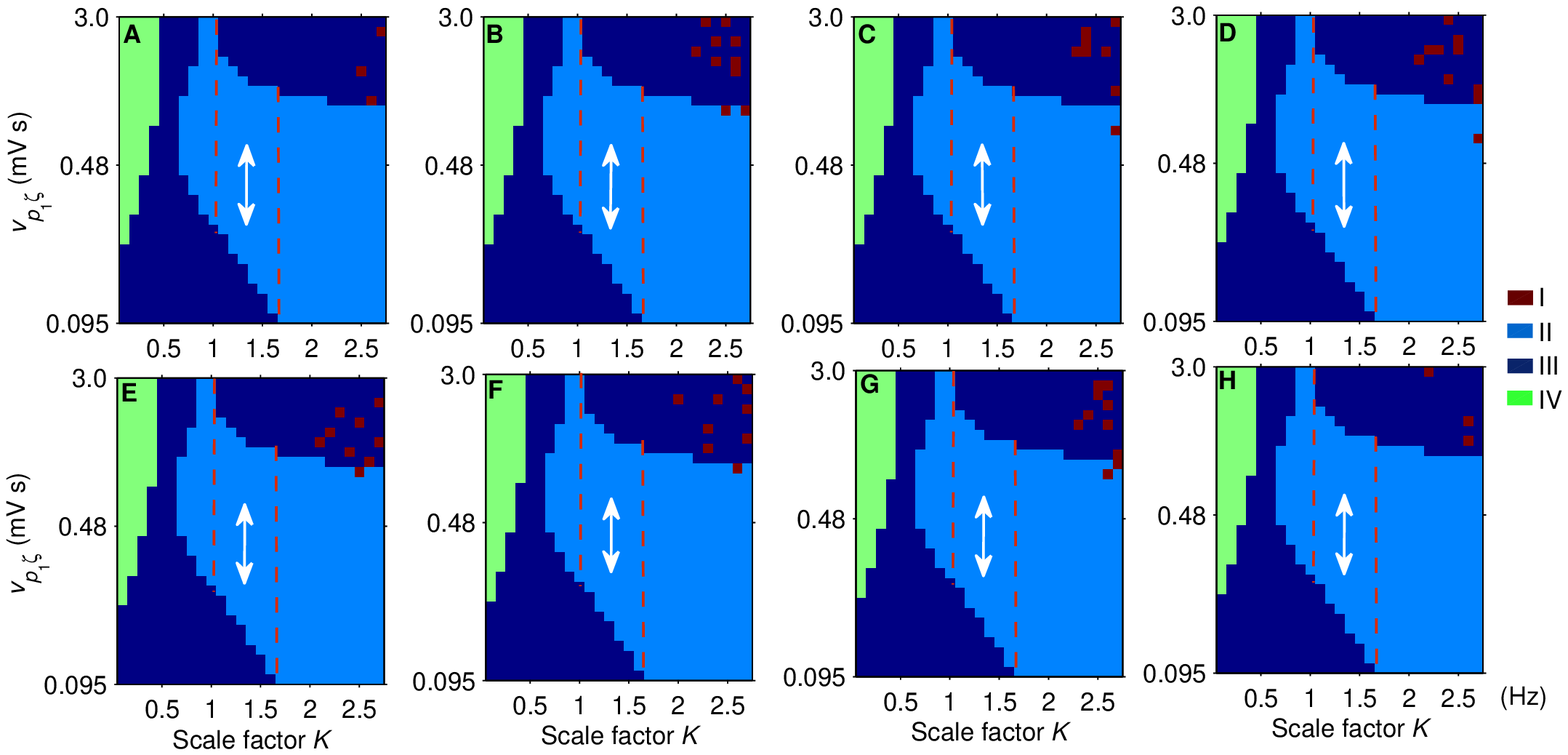}
\\
\textbf{[Figure S3 legend]}
\label{fig:Fig.S3}
\end{figure}
\end{landscape}

\end{document}